\documentclass[%
 aip,
 amsmath,amssymb,
 reprint,%
]{revtex4-1}

\usepackage[dvipsnames]{xcolor}
\usepackage{graphicx}
\usepackage{dcolumn}
\usepackage{bm}

\usepackage[utf8]{inputenc}
\usepackage[T1]{fontenc}
\usepackage{mathptmx}
\usepackage{etoolbox}
\usepackage{xspace}
\usepackage[normalem]{ulem}

\newcommand{\OpenFOAM}{OpenFOAM\textsuperscript{\tiny\textregistered}\xspace}
\newcommand{\Wi}{$\mathrm{Wi}$\xspace}

\newcommand{\Karman}{K\'arm\'an\xspace}

\newcommand{\mean}[1]{\overline{#1}}

\newcommand{\rev}[1]{{\color{NavyBlue}#1}}
\renewcommand{\rev}[1]{{\color{Black}#1}}

\makeatletter
\def\@email#1#2{%
 \endgroup
 \patchcmd{\titleblock@produce}
  {\frontmatter@RRAPformat}
  {\frontmatter@RRAPformat{\produce@RRAP{*#1\href{mailto:#2}{#2}}}\frontmatter@RRAPformat}
  {}{}
}%
\makeatother
\begin{document}

\preprint{POF21-AR-05587}

\title[]{Characterizing elastic turbulence in the three-dimensional
von \Karman swirling flow using the \mbox{Oldroyd-B} model}
\author{Reinier van Buel}
\author{Holger Stark}%
\affiliation{%
Technische Universit{\"a}t Berlin,
{Institute of Theoretical Physics,}
Hardenbergstrasse 36, 10623 Berlin, 
{Germany}
}
 \email{r.vanbuel@tu-berlin.de}

\date{\today}

\begin{abstract}
We present the full three-dimensional numerical investigation of the von \Karman swirling flow between two parallel plates using the \mbox{Oldroyd-B} model and characterize the onset and development of elastic turbulence.
We quantify the flow state with the secondary-flow strength, a measure of the average strength of the velocity fluctuations, and then define an order parameter as the time average of the secondary-flow strength.
The order parameter displays a subcritical transition from the
laminar to a bistable flow that switches between \rev{weakly chaotic flow} and elastic turbulence.
The transition to the bistable flow occurs at the critical Weissenberg number $\mathrm{Wi_c}=12$.
Above $\mathrm{Wi_c}$, in the elastic turbulent state, we observe a strong increase in velocity fluctuations and flow resistance, 
which we define as the total work performed on the fluid. Upon starting simulations in the turbulent state and subsequently lowering 
$\mathrm{Wi}$ below its critical value, we observe hysteretic behavior in the order parameter and the flow resistance, which is a common feature of a subcritical transition. Hysteresis has also been found in experiments. Additionally, we find power-law scaling in the spatial 
and temporal power spectra of the velocity fluctuations, characteristic for elastic turbulence.
The maximum values of the power-law exponents in our simulations are
$\alpha_t= 3.41$ for the temporal exponent and $\alpha_s= 3.17$ for the spatial exponent, which are remarkably close to the values obtained in experiments.
\end{abstract}


\maketitle

\section{Introduction}

Viscoelastic fluids, such as dilute polymer solutions, exhibit elastic instabilities and transitions from laminar to steady and unsteady flows, even at very small Reynolds numbers \cite{groisman2000,groisman2004elastic,schiamberg2006transitional,arratia2006elastic}.
The most prominent example of an unsteady flow state is called elastic turbulence, which was classified as turbulent over two decades ago, since it bears many similarities to inertial turbulence observed in Newtonian fluids at high Reynolds numbers \cite{groisman2000}.
Elastic turbulence exhibits increased velocity fluctuations with a power-law dependence of the power spectrum and an increase in the flow resistance of the polymer solution \cite{groisman2000,groisman2004elastic}.
Moreover, the nonlinear and time-dependent properties of viscoelastic fluids and especially elastic turbulence can be employed to increase heat and mass transport in liquids at the micron scale \cite{groisman2001efficient,groisman2004elastic,burghelea2007elastic,thomases2009transition,thomases2011stokesian}.
In Newtonian fluids this is extremely challenging since transport at the micron scale is dominated by diffusion.
Thus, viscoelastic fluids have promising properties, which are especially useful in microfluidic applications, such as lab-on-a-chip devices\ \cite{squires2005microfluidics}.
\rev{Moreover, a recent review provides perspectives and a roadmap for viscoelastic flow instabilities and elastic turbulence \cite{datta2021perspectives}.
}

\rev{
The instability of the regular base flow is driven by elastic stresses, which are generated by polymers stretched in velocity gradients. In flows with curvilinear streamlines and shear rates constant in time, the onset of the elastic instability is determined by the Weissenberg number Wi, the product of the polymer relaxation time and a characteristic shear rate. 
Beyond a critical Weissenberg number, viscoelastic fluids can undergo a transition from laminar flow to steady symmetry-breaking flows, for example observed in a cross-slot geometry {\cite{arratia2006elastic,sousa2018purely,poole2007purely}}, or to 
{aperiodic flows}
for example, in 
{Taylor-Couette flow \cite{larson1990purely}, von \Karman swirling flow \cite{mckinley1991observations,byars1994spiral}, serpentine channel or Dean flow \cite{ducloue2019secondary}, cone-and-plate flow \cite{mckinley1991observations}, cross-channel flow \cite{arratia2006elastic,sousa2018purely}, and lid-driven cavity flows \cite{pakdel1996elastic}.
Eventually, at high enough Weissenberg numbers, wall-bounded shear flows display elastic turbulence,
which has been well documented in experiments. Elastic turbulence has been observed in several of the geometries, already mentioned, such as Taylor-Couette flow \cite{groisman2004elastic}, von \Karman swirling flow \cite{groisman2000,groisman2001efficient,groisman2001stretching,groisman2004elastic,burghelea2007elastic,jun2009power,jun2017polymer,schiamberg2006transitional}, serpentine channel or Dean flow \cite{groisman2004elastic,soulies2017characterisation} and cross-channel flow \cite{sousa2018purely}.}
Moreover, temporal or spatial modulations of the shear rate can be used to control the onset of the elastic instability {and elastic turbulence \cite{vanBuel2020active,walkama2020disorder}.} 
}

A determining characteristic of elastic turbulence is the power-law decay of the spatial and temporal velocity power spectra, where the spatial scaling exponent 
has been demonstrated to be larger than three 
\cite{fouxon2003spectra}.
In contrast, the temporal power spectrum and its scaling exponent are more easily accessible in experiments. Now, Taylor's hypothesis for inertial turbulence in Newtonian fluids, which states that spatial and temporal exponent are equal\ \cite{taylor1938spectrum}, has also been suggested to be valid for elastic turbulence \cite{groisman2004elastic,sousa2018purely}.
Both exponents were reported in experiments \cite{groisman2000,groisman2001efficient,groisman2004elastic,burghelea2007elastic} and simulations \cite{berti2008two,berti2010elastic}.
Thereby, experimental work showed that Taylor's hypothesis can only be applied reasonably in regions where the mean flow velocity is large\cite{burghelea2005validity,burghelea2007elastic}. 
In our own numerical work on the two-dimensional Taylor-Couette flow, we found differences between the exponents for all flow strengths, while the overall behavior as a function of radial position was similar \cite{vanBuel2018elastic}.
Recently, scaling relations between the exponents of the power-law decays of several quantities, such as the elastic energy, pressure and torque fluctuations have been predicted \cite{steinbergscaling}. 
Furthermore, in numerical work an important observation is that the observed scaling in the velocity spectrum does not depend on the chosen polymer model \rev{\cite{steinbergscaling,berti2008two,berti2010elastic,vanBuel2018elastic,steinberg2021elastic}} and agrees well with the experimental values \cite{groisman2000,groisman2001efficient,groisman2004elastic,burghelea2007elastic,jun2017polymer,varshney2018drag,jun2009power}.

Detailed experiments on the parallel plate geometry at low Reynolds numbers showed
transitions to periodic, aperiodic, and turbulent flows \cite{groisman2000,groisman2004elastic,burghelea2007elastic,mckinley1991observations,schiamberg2006transitional}. 
Moreover, Schiamberg \textit{et al.} demonstrated a transitional pathway from the stable base flow at low Weissenberg numbers to elastic turbulence at higher Weissenberg numbers \cite{schiamberg2006transitional}. 
Upon increasing the Weissenberg number, they found several static and dynamic flow states between base and turbulent flow.
The static flow state comprises one or several axisymmetric annular disturbances of the base flow, and the dynamic states include a non-axisymmetric spiral wave, a superposition of multiple competing non-axisymmetric spiral waves that form at all radial locations, and a periodic pattern in time of spiral waves traveling radially outwards.
A subcritical transition from the stable base flow to chaotic flow is observed in most experiments \cite{groisman2004elastic,burghelea2007elastic,mckinley1991observations}, whereas Schiamberg \textit{et al.} find continuous transitions between the different flow states \cite{schiamberg2006transitional}.
\rev{
In summary, the transitional path towards elastic turbulence depends on different parameters such as
the gap-to-plate-radius ratio, the Reynolds number, the lateral boundary conditions, and the polymer concentration. However, the properties of elastic turbulence remain the same.
}

A subcritical transition was also reported in theoretical work by Walsh \cite{walsh1987flow}, who analyzed fluid flow 
between infinitely extended parallel plates rotating relative to each other based on the upper-convective Maxwell model. The transition occurred at the same critical Weissenberg number as the one found by 
Phan-Tien using linear stability analysis \cite{phan1983coaxial}. 
Both approaches use a similarity transformation such that the velocity components depend linearly on the radial position and are otherwise only functions of the height.
In contrast, instabilities observed in experiments develop from localized flow disturbances \cite{mckinley1991observations,byars1994spiral}.
They start at a specific radius and then travel inward or outward. The similarity transformation is unable to  capture such an instability.
An alternative approach for a linear stability analysis realized by {\"O}ztekin \textit{et al.} uses a large radius approximation and concentrates the disturbances around a critical radius \cite{oztekin1993instability}.
\rev{In experiments it hints to the radius where the shear rate is largest (often close to the plate radius), which is where the instability starts.}
While the results agree well with the experiments of McKinley \textit{et al.} \cite{mckinley1991observations}, a drawback of this method is that it disregards spatially coupled solutions and edge effects due to the finite extent of the rotating plates.
To fully address the problem, solutions to the full three-dimensional eigenvalue problem are required. However, so far such investigations have not been presented yet.

\rev{
Direct numerical simulations have widely been employed to find solutions to the nonlinear equations governing viscoelastic fluids, {specifically at low Reynolds numbers}. 
{Numerical work has demonstrated elastic instabilities in similar geometries as the ones explored in the experiments but limited to two-dimensional flows. 
This includes the cross-slot geometry \cite{davoodi2019control,poole2007purely,xi2009mechanism}, sudden-expansion flow in widening channels \cite{poole2007plane}, serpentine channels \cite{poole2013viscoelastic},
two cylinders confined in a channel \cite{kumar2021elastic},
two lateral side-by-side cylinders in a channel \cite{hopkins2021tristability},
and the Taylor-Couette geometry \cite{davoodi2018secondary}.
{Recent numerical work has demonstrated the elastic instability in the von \Karman swirling flow \cite{khambhampati2020numerical}.}
Furthermore, simulations have identified elastic turbulence in two dimensions in the}
cross-slot geometry \cite{Canossi2020elastic}
and the Taylor-Couette flow \cite{vanBuel2018elastic,vanBuel2020active}.
Additionally, articles address unbounded flows with 
{cellular} forcing, {where they observe elastic instabilities} \cite{gutierrez2019proper,thomases2011stokesian} 
{and elastic turbulence \cite{gupta2019effect,berti2008two,berti2010elastic},
or employ a {shell} model \cite{ray2016elastic}.}
Furthermore, numerical calculations using the pseudo-spectral method {in three dimensions} have shown the elastic instability in 
Taylor-Couette flow \cite{thomas2006pattern}. 
}

\rev{In this work we provide detailed direct numerical simulations of the fully three-dimensional von \Karman swirling flow between two parallel plates.
Our goal is to explore the general features of the elastic instability and elastic turbulence and not to model one specific experiment. Therefore, we follow other studies and as a starting point use the \mbox{Oldroyd-B} model \cite{shaqfeh2021oldroyd,beris2021continuum,thompson2021reynolds,castillo2022understanding}, which describes an idealized viscoelastic fluid with a minimum number of free parameters.
For example, it does not include a shear-dependent viscosity similar to Boger fluids \cite{boger1977highly}.}
We first analyze the stability of the base flow through a linear stability analysis following Ref.~\onlinecite{oztekin1993instability} 
and present neutral stability curves. 
They show a non-axisymmetric mode with three-fold symmetry as the most unstable one but with other modes close by. Then, we investigate the onset and development of the instability in our simulations and find an unstable non-axisymmetric mode with four-fold symmetry driving the instability towards \rev{weakly chaotic flow}.
Analyzing the velocity fluctuations close to the 
\rev{subcritical}
transition
\rev{at the critical Weissenberg number $\mathrm{Wi_c}$,}
we identify a bistable flow state, which switches between \rev{weakly chaotic} and turbulent 
flow, and we quantify them by plotting an appropriate order parameter versus the Weissenberg number.
\rev{The subcritical transition gives rise to hysteretic behavior. Below $\mathrm{Wi_c}$ only the weakly chaotic state remains,
which shows, on average, the signature of the three-fold symmetric mode predicted by linear stability analysis.}
A thorough analysis of the flow resistance, which we define as the total work
performed on the fluid, reveals a sharp increase at the transition to elastic turbulence due to the elastic work performed 
at the \rev{lateral sides} of the swirling fluid.
Finally, we thoroughly analyze spatial and temporal velocity power spectra detailing the turbulent nature of the flow.

The remainder of our article is structured as follows.
In section \ref{sec:theory}, we discuss the governing equations of the \mbox{Oldroyd-B} model, our computational method including the simulation parameters, and we briefly explain our implementation of the linear stability analysis. 
We derive the equations for the linear stability analysis and validate our results in Appendix\ \ref{app1}.
In section \ref{sec:lsa}, we present the results from our linear stability analysis. 
We display the onset of the flow instability observed in our simulations in section \ref{sec:onset}.
In section \ref{sec:ET}, we characterize the bistable flow through the secondary-flow strength and introduce an order parameter. 
Results of the flow resistance and the observed hysteretic behavior in the flow are also presented. Furthermore, in 
section \ref{sec:PSD}, we demonstrate power-law scaling of the spatial and temporal velocity spectra and analyze the scaling 
exponents.
\rev{In section \ref{sec.discussion}, we discuss our results in relation to experiments and, finally,}
we conclude in section \ref{sec:conclusion}.

\section{Theory and Methods}
\label{sec:theory}

\begin{figure}
\centering
\includegraphics[width=0.45\textwidth]{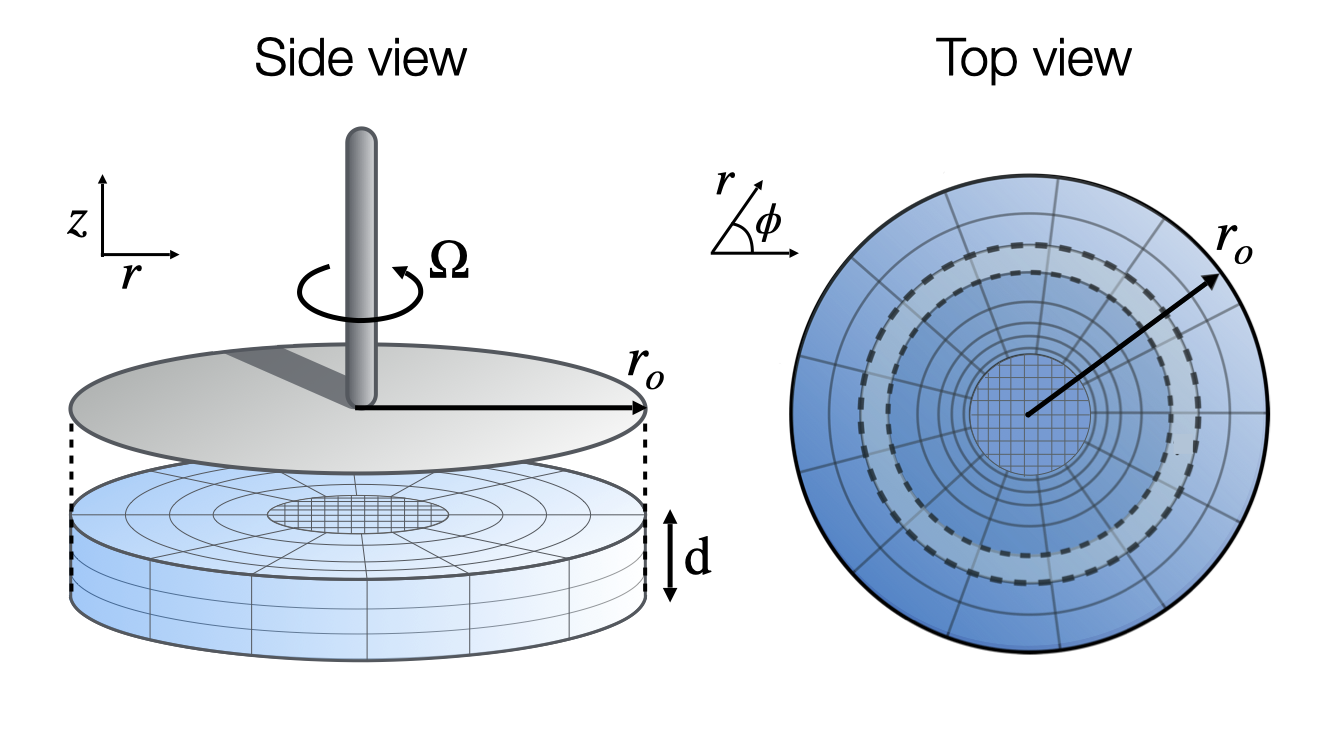}
\caption{
Schematic of the three-dimensional parallel plate geometry for simulating von \Karman swirling flow and the spokes-wheel mesh used in \OpenFOAM \, with $N_r = 100$ cells in the radial direction, $N_{\phi} = 120$ cells in the angular direction, and $N_{z} = 10$ cells in the z-direction. 
Note, $30\times30$ cells are distributed on a square lattice in the inner cylinder block. $\Omega$ is the angular velocity of the top plate with radius $r_o$, the bottom plate  is fixed, and $d$ is the distance between the plates.
\rev{For the angular average along the azimuthal angle $\phi$, we take the field values of cells with a fixed radius. An example is illustrated by the lighter shaded area surrounded by dashed lines.
For the radial average along the radial direction, each contribution
from a cell is weighted by the volume of the specific cell.
}
}
\label{fig:schematic}
\end{figure}

We consider an incompressible viscoelastic fluid in the geometry of
a three-dimensional von \Karman swirling flow, where the viscoelastic fluid is constrained between two parallel plates with radius $r_o$ and the upper plate is rotating with angular velocity $\Omega$.
A schematic of our set-up can be seen in Fig.~\ref{fig:schematic}.
\rev{We use the characteristic length $r_o$
and the characteristic velocity $r_o\Omega$} 
to define the Reynolds number \mbox{$\mathrm{Re} = \rho \Omega r_o^2 / \eta_s$}, where $\eta_s$ is the solvent shear viscosity. In the following, we set the Reynolds number $\mathrm{Re} \ll 1$, since we are interested in viscoelastic fluid flow at small scales, such as in microfluidic settings.
\rev{In the following we always use the period of the plate rotation, $2\pi/\Omega$, to rescale time and frequency.}

We calculate the dynamics of the flow field $\bm{u}(\bm{r},t)$, where $\bm{r}$ denotes the position and $t$ the time, using the generalized Navier-Stokes equation for an incompressible \mbox{Oldroyd-B} fluid. 
\rev{ Neglecting the nonlinear inertial term to speed up the calculations, we have}
\begin{align}
\label{eq:NS} 
\nabla \cdot \bm{u} &= 0 \, , \\
\rho 
\frac{\partial \bm{u}}{\partial t} 
&= 
-\nabla p
+\eta_s \nabla^2 \bm{u} +  {\nabla \bm{\tau}} \, .
\end{align}
Here, $\rho$ is the density of the solvent, $p$ is the pressure, $\eta_s$ the solvent shear viscosity, and ${\nabla \bm{\tau}}$ denotes the divergence of the stress tensor $\bm{\tau}$.
It describes the viscoelastic stresses, for example, due to dissolved polymers in polymer solutions.
Concretely, we choose the constitutive relation of the \mbox{Oldroyd-B} model \cite{oldroyd1958non,oldroyd1950formulation} to model the 
viscoelastic stresses, which reads
\begin{equation}
\label{eq:Oldroyd}
\bm{\tau} +\lambda \overset{\nabla}{\bm{\tau}} = 
\eta_p \left( \bm L  + \bm L ^\mathrm{T} \right) \, .
\end{equation}
Here, $\bm L = \nabla \otimes \bm{u}$ is the velocity gradient tensor, $\eta_p$ is the polymeric shear viscosity and $\lambda$ a characteristic relaxation time of the dissolved polymers.
Lastly, $\overset{\nabla}{\bm{\tau}}$ denotes the upper convective derivative of the stress tensor defined as
\begin{equation}
\overset{\nabla}{\bm{\tau}} = \frac{\partial  \bm{\tau}}{\partial t} + \bm{u}\cdot \nabla\bm{\tau} - \bm L  \bm{\tau} - \bm{\tau}  \bm L ^\mathrm{T}  \, .
\end{equation}

The governing equations {(\ref{eq:NS})-(\ref{eq:Oldroyd})}
can be written in dimensionless form with three relevant parameters.
Besides the Reynolds number $\mathrm{Re}$, one has the ratio $\beta = \eta_p/\eta_s$ of the polymeric shear viscosity to the solvent shear viscosity and the Weissenberg number $\mathrm{Wi} =  \lambda \dot{\gamma}$, where $\dot{\gamma}$ is the characteristic shear rate.
For the parallel plate geometry the characteristic shear rate is the ratio of the angular velocity of the rotating plate to the gap width, {$ \dot{\gamma} = \Omega {r_o}/{d}$}, and we have $\mathrm{Wi} = \lambda  \Omega {r_o}/{d}$.

\subsection{Computational method}
Following our previous work \cite{vanBuel2018elastic},
Eqs.~(\ref{eq:NS})-(\ref{eq:Oldroyd}) are solved using the open-source program \OpenFOAM \cite{weller1998tensorial}, which is a finite-volume solver for computational fluid dynamics simulations on polyhedral grids. 
We adopt a specialized solver for viscoelastic {fluids}
called rheoTool \cite{favero2010}, which is implemented in \OpenFOAM.
The rheoTool solver has been tested for accuracy in benchmark flows and it has been shown to have second-order accuracy in space and time \cite{pimenta2017}.

Numerical evaluations of viscoelastic fluid flow can be unstable when the conformation tensor loses its positive definiteness due to numerical discretization errors. Regions near stagnation points or regions with strong deformation rates are sensitive to numerical instabilities and the effect is especially significant at high Weissenberg numbers \cite{fattal2005}.
Stability of the numerical flow field can be increased by introducing the log-conformation tensor approach \cite{fattal2004}, which is based on taking the logarithm of the conformation tensor and which is implemented in the rheoTool solver.
The positive definite conformation tensor $\bm{\mathrm{C}}$ is related to the polymeric stress tensor by 
\begin{equation}
\label{eq:afine}
\bm{\tau} = \frac{\eta_p}{\lambda} \left(\bm{\mathrm{C}} - \bm{\mathrm{I}}\right) \, ,
\end{equation}
where $\bm{\mathrm{I}}$ is the identity tensor.
Now, setting $\bm{\mathrm{\Theta}} = \ln \bm{\mathrm{C}}$, the constitutive relation (\ref{eq:Oldroyd}) is transformed {to} a dynamic equation for $\bm{\mathrm{\Theta}}$, which is then numerically evaluated. More details of the method are given in Ref.~\onlinecite{fattal2004}. 
Evolving $\bm{\mathrm{\Theta}}$ in time and then transforming back to the conformation tensor $\bm{\mathrm{C}}$ and stress tensor  $\bm{\tau}$ gives enhanced stability \cite{pimenta2017}.
However, the error in the conformation tensor $\bm{\mathrm{C}}$ increases, which is mitigated by setting a small time step in the numerical evaluation and by using a mesh refinement towards the inner cylinder, where discretization errors of the employed mesh increase. 
In our numerical calculations we further use a biconjugate gradient solver combined with a diagonal incomplete LU preconditioner 
(DILUP-BiCG) to solve for the components of the polymeric stress tensor and a conjugate gradient solver coupled to a diagonal incomplete Cholesky preconditioner (DIC-PCG) to solve for the velocity and pressure fields.
They are available in \OpenFOAM following the work of Ref.\ \onlinecite{pimenta2017}.

In the schematic of the parallel plate geometry in Fig.~\ref{fig:schematic} two types of lattices can be seen, a radial and a square lattice.
The square lattice is implemented to remove the radial singularity in the middle of the cylinder. It consists of $\left(N_\phi/4 \right)^2$ cells and is connected to the radial lattice at $r=0.1\, r_o$. 
At the inner radius of the radial lattice we choose a finer radial mesh compared to the outer cylinder such that the radial width $\Delta r$ of the grid cells is ${2.0}\mathrm{\cdot 10^{-3}}\, r_o$ at $0.1\, r_o$ and increases to $1.9 \cdot 10^{-2} \,r_o$ at $r_o$.

At the lower and upper bounding plates we choose the no-slip boundary condition for the velocity field, a zero gradient for the pressure field, and extrapolate the gradient for the polymeric stress field to zero.
\rev{Hence, the velocity at the lower plate is $\bm{u} = \bm{0}$ and at the upper plate $\bm{u} = r \Omega \bm{e}_\phi$.}
Importantly, at the sides of the cylindrical cell we set the normal component of the velocity field to zero ($u_r = 0$) and apply the same boundary conditions for pressure and polymeric stress field as before.
The simulations start with the viscoelastic fluid at rest, where pressure, flow, and stress fields are uniformly zero.

The following geometric parameters are chosen from the viewpoint of a microfluidic setting such that the Reynolds number is low.
As required by \OpenFOAM{}, we give all parameters in {Si} units.
The outer radius of the geometry is set to $r_o = 10^{-5}\,\mathrm{m}$, the height to $d=0.1\, r_o$, and the rotational velocity is $\Omega = 0.38 \,\mathrm{s^{-1}}$.
We adjust the Weissenberg number by varying the polymeric relaxation time $\lambda \,\in [0,\rev{9.9\, \mathrm{s} }] 
$.
We set the polymeric shear viscosity to $\eta_p=0.0015 \, {\mathrm{kg/ms}}$, the solvent shear viscosity to 
\mbox{$\eta_s=0.001  \, {\mathrm{kg/ms}}$}, and the density to $\rho=1000 \,\mathrm{kg/m^3}$.
The ratio of the polymeric to the solvent viscosity is then $\beta={\eta_p}/{\eta_s}=3/2$. 
For the Reynolds number we obtain $\mathrm{Re} = \rho \Omega r_o^2 / \eta_s = 0.038$.
\rev{We note to obtain a small Re in experiments large polymeric and solvent viscosities are commonly employed. 
Having in mind microfluidic settings, in our work we choose a small characteristic length, which is equivalent to studying polymeric fluids at larger spatial dimensions 
with a larger solvent and polymeric viscosity.
The fluid flow is simulated up to ${800}\,\mathrm{s}$ {(about $50$ to $235 \lambda$)} with a time step  $\delta t = 10^{-5}\,\mathrm{s}$ {($10^{-5}$ to $ 10^{-6} \lambda$)}.
We extract the velocity, pressure, and stress fields every 5000 time steps {(about $0.005$ to $ 0.02 \lambda$)}. The time step and extraction interval are very small compared to the relaxation time.}

\subsection{Linear stability analysis}
\label{subsec.linear_analysis}

To investigate the stability of the von \Karman swirling flow of the \mbox{Oldroyd-B} fluid, we perform a linear stability analysis following the works of Refs.~\rev{
\onlinecite{avgousti1993non} and \onlinecite{oztekin1993instability}}.
A perturbation is superimposed on the base flow solution for which we choose a product ansatz in the three cylindrical coordinates $r$, $\phi$, and $z$. Concretely, following Ref.~\onlinecite{oztekin1993instability},
we analyze the stability of the base flow ($ \bm{\tau} ^0, \bm{u}^0,p^0$) to radially localized small-amplitude disturbances of the form
\begin{align}
\label{eq:linstab1}
\tilde{\bm{\tau}}  &=  \bm{\tau} ^0 + \mathcal{R} \left[\bm{\tau} (z)\exp(i \alpha r + i m \phi + \sigma t)\right] \, , \\
\label{eq:linstab2}
\tilde{\bm{u}} &=   \bm{u}^0 + \mathcal{R}\left[\bm{u}(z)\exp(i \alpha r + i m \phi + \sigma t)\right] \, , \\
\label{eq:linstab3}
\tilde p &= p^0 +  \mathcal{R}\left[ p(z)\exp(i \alpha r + i m \phi + \sigma t) \right] \, ,
\end{align}
where $\mathcal{R}[.]$ represents the real part, $\bm{\tau} , \bm{u}, p$ are small complex functions of the $z$ coordinate, $\alpha$ is the real-valued radial wave number, $m$ is {the} azimuthal wave number, and $\sigma$ is the complex frequency, which we determine from the eigenvalue problem.
The azimuthal wave number $m$ is an integer, which can be positive or negative for non-axisymmetric disturbances, while $m=0$ corresponds to axisymmetric disturbances.

For the linear stability analysis we take the base flow 
solution 
{of}
the governing equations 
{(\ref{eq:NS})-(\ref{eq:Oldroyd})} 
with the boun\-da\-ry condition
\begin{align}
\label{eq:ubase}
    \bm{u}(r,\phi,0) = \bm{0} \enspace ,
\qquad \bm{u}(r,\phi, d ) = 
{r\Omega \bm{e}_\phi} \, ,
\end{align}
{where $\bm{e}_\phi$ is the unit vector in the azimuthal direction.}
Neglecting fluid inertia 
{($\text{Re} \approx 0$),} 
the steady-state 
velocity field
{for infinitely extended parallel plates}
is purely 
{azimuthal} 
and given by
\begin{align}
\label{eq:baseflow}
    \bm{u}^0 = \frac{r\Omega z }{ d } \bm{e}_\phi \, .
\end{align}
The corresponding base stress and pressure fields read
\begin{align}
\label{eq:taubase}
    \tau_{rr}^0 = \tau_{r\phi}^0 = \tau_{r z}^0 = \tau_{zz}^0 = 0; \qquad \tau_{\phi z}^0 = \eta_p \frac{r\Omega}{d} ; \\ \tau_{\phi\phi}^0 = 2 \eta_p \lambda \left(\frac{r\Omega}{d}\right)^2 ; \qquad
    p^0 = -\eta_p \lambda \left(\frac{r\Omega}{d}\right)^2 \, .
\end{align}

After substituting {Eqs.\ (\ref{eq:linstab1})-(\ref{eq:linstab3})} in the governing equations (\ref{eq:NS})-(\ref{eq:Oldroyd}),
using the fact that the base flow solves these equations,
and disregarding all terms that are nonlinear in the disturbance amplitude, we obtain a set of 10 linear differential equations 
(see Appendix \ref{app1}).
One can show that the linear stability analysis can be reduced to an ODE system of the form $\bm{\psi}' = F(z,\bm{\psi})$ with 6 independent variables, which we express with the vector 
\begin{align}
    \bm{\psi} = [u_r(z),u_\phi(z), p(z) , u_r'(z),u_\phi'(z), p'(z) ]^\mathrm{T} \, ,
\end{align}
where the prime indicates the derivative $g'=dg/dz$ with respect to the spatial coordinate $z$. The remaining variables, such as the components of the polymeric stress tensor, are then obtained from the independent variables.

The boundary conditions of the velocity perturbation are set to
\begin{align}
    \bm{u}(r,\phi,z) = 
    {\bm{0}}
    \qquad \text{at $z = 0$ and $z = 1$} \, .
\end{align}
Moreover, we take the radially localized perturbation to occur at the edge of the rotating cylinder, $r=r_o$, where the shear rate is the highest
and where we expect the perturbation to be strongest. Following Ref.\ \onlinecite{oztekin1993instability} this also means that in the set of linear equations we replace $r$ by $r_o$.
We numerically solve the linear stability problem and identify the eigenvalue $\sigma$ in an iterative manner using a collocation method, which discretizes the perturbation functions in the $z$ direction, with the computer algorithm \textit{solve\_bvp} available in the SciPy package \cite{2020SciPy-NMeth}.

We have validated our numerical algorithm against the results of Ref.\ \onlinecite{oztekin1993instability} and find good agreement with our findings as we demonstrate in Appendix \ref{app1}. The validation is essential given the large number of terms involved in the calculations.
In section\ \ref{sec:lsa} we perform the linear-stability analysis for our parameter setting.

\section{Results}

In this section we describe our results. First, we analyze the stability of the base flow through linear stability analysis. Second, we display the initial evolution of an non-axisymmetric instability observed in our simulations before turbulent flow fully sets in.
Then, we characterize the flow state above the critical Weissenberg number, where we observe turbulent and \rev{weakly chaotic flow}.
We further characterize both flow states through the flow resistance, the amount of work performed on the fluid surface. Finally, we analyze spatial and temporal power spectra of the velocity and stress fields.

\begin{figure}[b]
\centering
\includegraphics[width=0.43\textwidth]{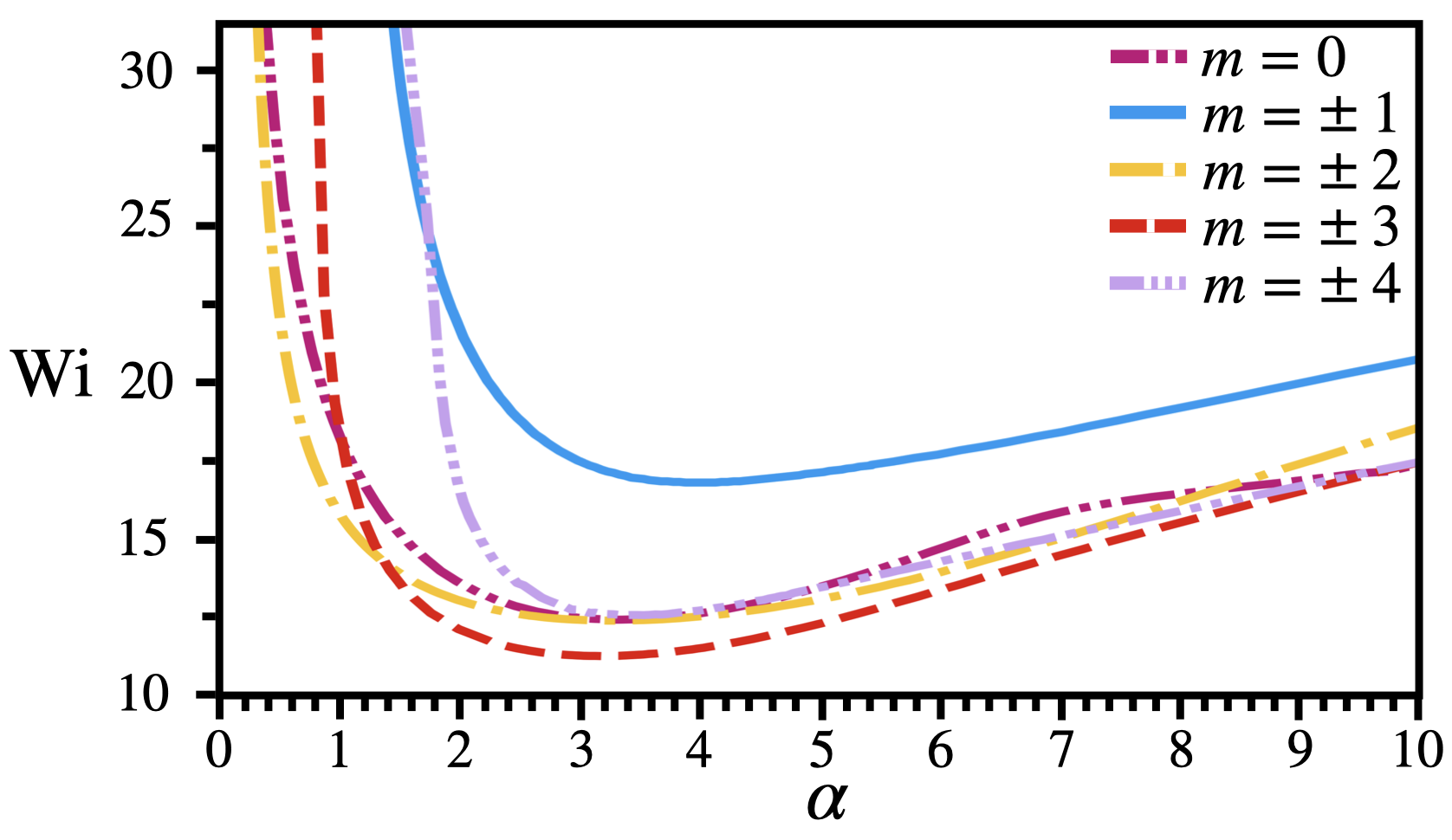}
    \caption{Neutral stability curves for the \mbox{Oldroyd-B} fluid in a von \Karman swirling flow 
    {in the plane Weissenberg number \Wi versus radial wave number $\alpha$}
       for the first nine azimuthal modes \mbox{($-4 \leq m \leq 4$)}. 
    {The stability curves of the different modes are indicated in the legend.}
The polymer to solvent viscosity ratio is set to 
\rev{$\beta = 3/2$} and the critical radius to $r_o=10\,d$.
}
  \label{fig:linstab}
\end{figure}

\begin{table}[t]
\caption{
The axisymmetric ($m=0$) and non-axisymmetric \mbox{($|m|=1,2,3$)} modes become unstable at the critical Weissenberg number 
$\mathrm{Wi_c^{LSA}}$. Their critical axial wave number $\alpha_c$ and the oscillation frequency $\sigma_\mathrm{im}$
from the imaginary part of $\sigma$ are indicated. The polymer to solvent viscosity ratio is set to $\beta = 3/2$ and the critical 
radius is $r_c = r_o = 10 \, d$.
}
\renewcommand{\arraystretch}{1.25}
\begin{ruledtabular}
\begin{tabular}{cccccc}
 { $|m|$ } & 0 & 1 & 2 & 3 & 4 \\ 
$\mathrm{Wi_c^{LSA}}$ & 12.42 & 17.00  & 12.389 & 11.255 &  12.574  \\
$\alpha_c$  & 3.30 & 3.50  & 3.30  & 3.25 & 3.50  \\ 
$\sigma_\mathrm{im}$ & -1.25 & -1.25  &  -0.375 & -0.345 & -0.345 \\ 
\end{tabular}
\end{ruledtabular}
\label{tab:linstab}
\end{table}
\renewcommand{\arraystretch}{1.0}

\subsection{Linear stability analysis}
\label{sec:lsa}

First, we investigate the stability of the von \Karman base flow (see Fig. \ref{fig:schematic}) using the methodology outlined in section\ \ref{subsec.linear_analysis}. We set the ratio of polymer to solvent viscosity to $\beta = 3/2$
and the critical radius to $r_o=10 \, d$.
The critical conditions of the linear stability equations for axisymmetric and 
non-axisymmetric instabilities depend on the Weissenberg number \Wi and on the 
radial wave number $\alpha$, which is a function of \Wi.
Hence, for each wave number the Weissenberg number where the first 
eigenmode becomes unstable (real part of eigenvalue $\sigma_\mathrm{re} >0$) is determined.
The neutral stability curves ($\sigma_\mathrm{re} = 0$) separating stable from unstable regions for the axisymmetric ($m=0$) and non-axisymmetric ($0 < |m| \leq 4$) modes are presented in the $\alpha$-\Wi plane in Fig. \ref{fig:linstab}.
The critical wave number $\alpha_c$ for each $m$ is then obtained from the absolute minimum of 
each curve.
However, the minima of all the curves are shallow and excluding $m=\pm 1$ they are close to each other, implying that bands of axial wave numbers become unstable when the 
Weissenberg number is slightly increased above the critical value $\mathrm{Wi}_c$.
The results of the most unstable axisymmetric and non-axisymmetric modes are presented in Table\ \ref{tab:linstab}.

First of all, we find that the von \Karman base flow becomes linearly unstable,
which first happens for non-axisymmetric disturbances with azimuthal wave numbers $m= \pm 3$.
The instability occurs at the critical radial wave number $\alpha_c = 3.25$ and critical Weissenberg number  
$\mathrm{Wi_c^{LSA}}=11.255$, where the superscript refers to the result of the linear stability analysis.
Moreover, the neutral stability curves for the axisymmetric mode and two 
non-axisymmetric modes ($|m|=2,4$) are very close to each other around the mi\-ni\-ma. This demonstrates that multiple modes become unstable when the Weissenberg number is slightly increased above the critical value of the $|m|=2$ modes.
We note that shallow minima are also observed in the neutral stability curves of Taylor-Couette flow at low Reynolds numbers in Ref.~\onlinecite{avgousti1993non}.

\begin{figure}
\centering
\includegraphics[width=0.45\textwidth]{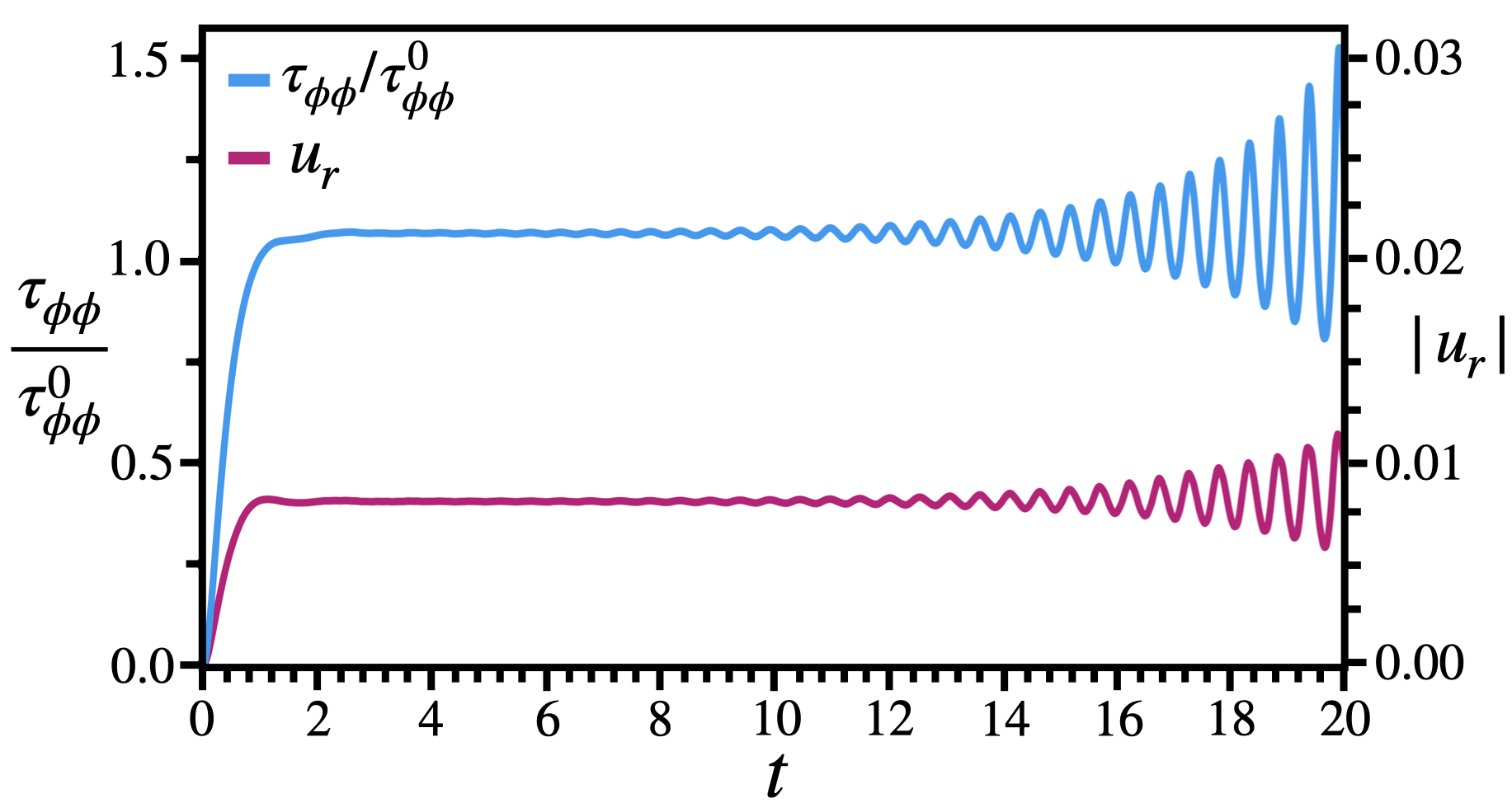}
    \caption{Azimuthal component of the stress tensor $\tau_{\phi\phi}$ normalized by the base flow $\tau_{\phi\phi}^0$ (left axis) and the magnitude of the radial velocity component $|u_r|$ (right axis) plotted versus time $t$ at $\mathrm{Wi}=12.008$
    {The components are recorded at position $(r,\phi,z) = (0.99\, r_o,0,d/2)$.}
    }
  \label{fig:initial}
\end{figure}

\begin{figure*}
\centering
\includegraphics[width=0.72\textwidth]{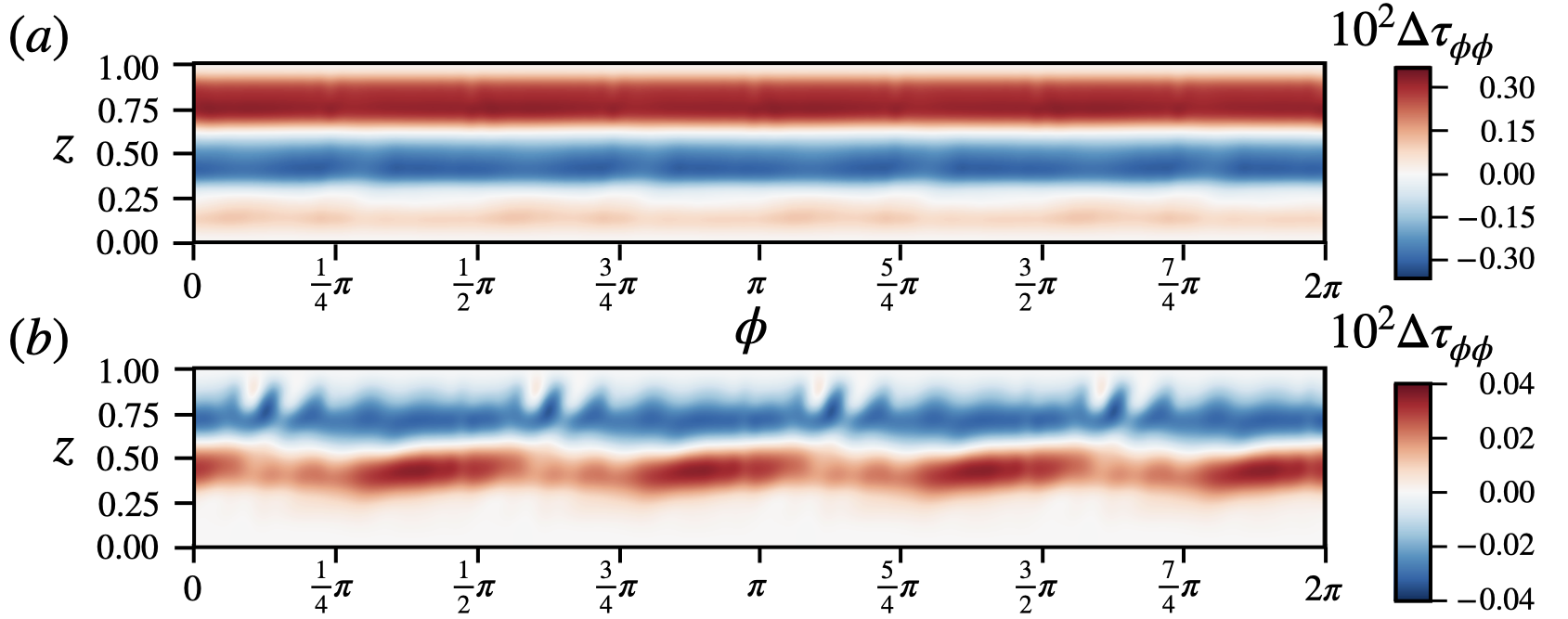}
\includegraphics[width=0.72\textwidth]{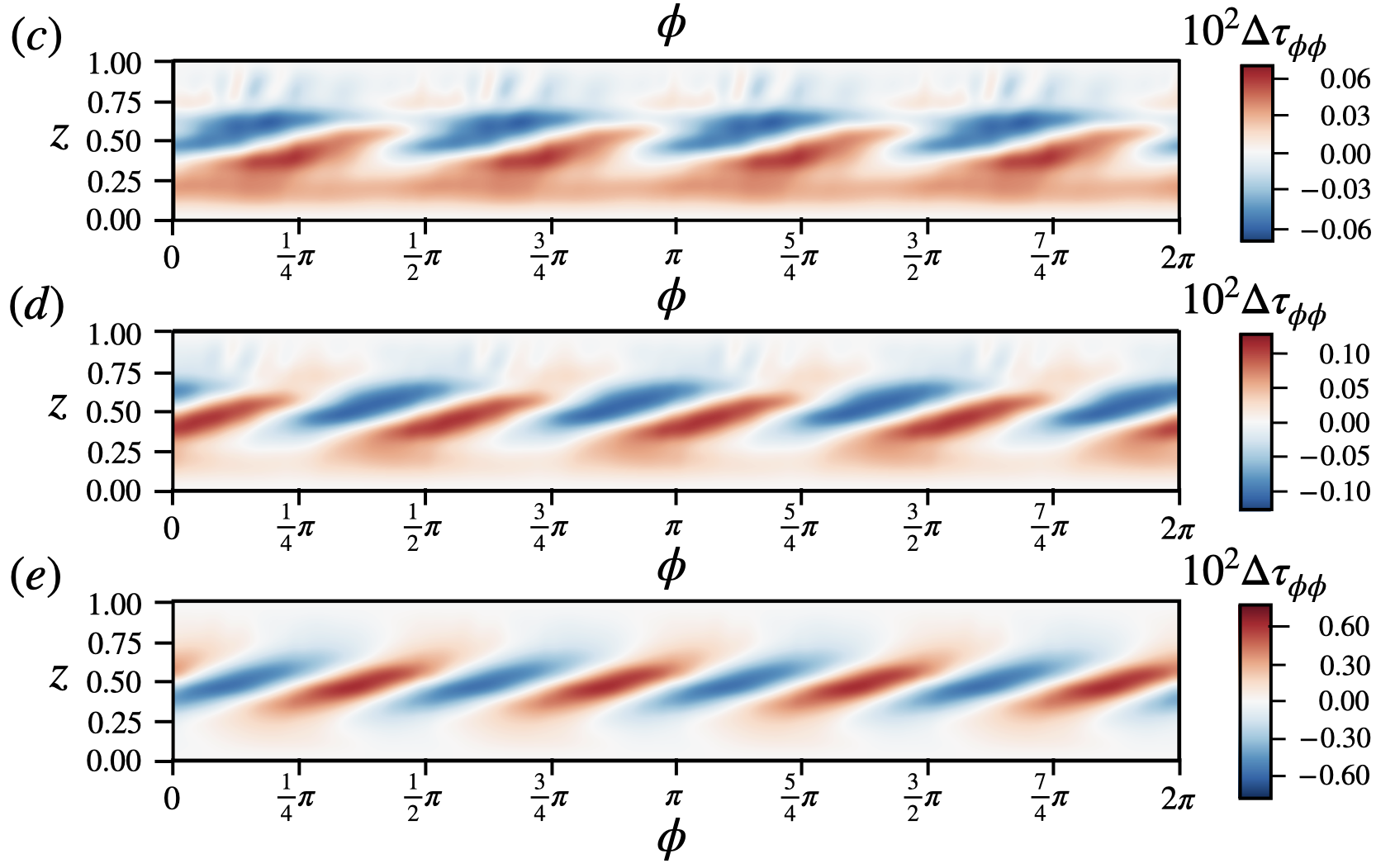}
\caption{(Multimedia view) Azimuthal stress-tensor component color-coded
in the $z-\phi$ plane near the outer radius $r=0.99\,r_o$ at $\mathrm{Wi}=12.008$.
{The component $\Delta \tau_{\phi\phi} = \tau_{\phi\phi}-\overline{\tau_{\phi\phi}}|_{t_s}^{t_e}$ is given relative to the mean value
between times $t_s=2$ and $t_e=3$.}
Snapshots of the stress field are taken at times: \mbox{(a) $t=1.4$}, \mbox{(b) $t=2.5$}, \mbox{(c) $t=7.6$}, \mbox{(d) $t=10.0$} and \mbox{(e) $t=15.0$}.
}
  \label{fig:Transition}
\end{figure*}

\subsection{Onset of the instability}
\label{sec:onset}

Now, we investigate the onset of the flow insta\-bi\-li\-ty occurring in the 
simulations of the \mbox{Oldroyd-B} fluid of the von \Karman swirling flow.
In contrast to the results from section\ \ref{sec:lsa}, we find a critical Weissenberg number around $\mathrm{Wi_c}\approx 12$ and observe, before the turbulent flow fully develops, an unstable non-axisymmetric mode with azimuthal wave numbers $m=\pm 4$, which we describe below.

All simulations start with the fluid at rest. During the initial state an axisymmetric disturbance flow occurs, which begins at the outer edge and then travels inwards\rev{, see for example Fig.~\ref{fig:disturbance}}. 
For low Weissenberg numbers $\mathrm{Wi} < \mathrm{Wi_c}$ it develops into a stable laminar flow.
However, at $\mathrm{Wi} = 12.008$ slightly above the critical value from our linear stability analysis, $\mathrm{Wi_c^{LSA}}=11.255$, we observe an unstable periodic disturbance flow.
The temporal evolution of the disturbance is illustrated in Fig. \ref{fig:initial}, where we plot the azimuthal component of the stress tensor $\tau_{\phi\phi}$ normalized by the base flow value $\tau_{\phi\phi}^0$ and the 
radial component of the velocity field $|u_r|$ at position $(r,\phi,z) = (0.99\, r_o,0,d/2)$. 
Both quantities oscillate with frequency $f=0.522$ and grow exponentially. 
Similar behavior occurs at $\mathrm{Wi}=12.16$ and $\mathrm{Wi}=12.92$.
Thus, above $\mathrm{Wi} > 12$ we observe a single unstable mode, which ultimately develops into a chaotic flow state.

Furthermore, we display the spatial and temporal appearance of the flow instability in Fig. \ref{fig:Transition} (Multimedia view), where we show the color-coded stress component $\tau_{\phi\phi}$ in the $z-\phi$ plane near the outer cylinder ($r=0.99 \,r_o$) in time and present snapshots at several instances.
First, Fig. \ref{fig:Transition} (a) shows the axisymmetric stress pattern,
which has a positive stress value at the upper and lower plate and a negative value in between.
The disturbance travels radially inward. At later time, in Fig.\ \ref{fig:Transition} (b) the line of zero stress close to the midplane between the two plates ($z=d/2$) becomes sinusoidal. Thus, a non-axisymmetric mode with four-fold symmetry de\-ve\-lops. Its amplitude grows further in time and the zero-stress line strongly deviates from the sinusoidal shape [see Fig. \ref{fig:Transition} (c)].
Ultimately, tilted regions of positive and negative azimuthal stress appear along nearly the entire height of the von \Karman geometry. They alternate in the azimuthal direction, as Figs.\ \ref{fig:Transition}(d) and (e) show. The amplitude of the pattern grows steadily in time and the whole pattern starts to slowly rotate about the vertical in the direction of the base flow until a chaotic flow pattern emerges.
The development of the instability is reminiscent of a Kelvin-Helmholtz instability observed in sheared Newtonian fluids {\cite{chandrasekhar2013hydrodynamic}}.

\begin{figure*}[t!]
\centering
\includegraphics[width=0.72\textwidth]{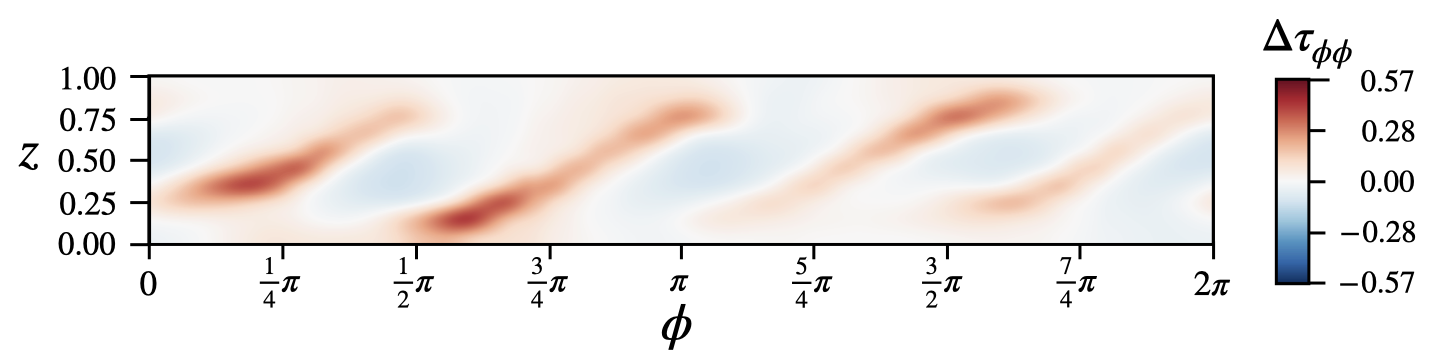}
\caption{(Multimedia view) Azimuthal stress-tensor component color-coded
in the $z-\phi$ plane near the outer radius $r=0.99\,r_o$ at $\mathrm{Wi}=12.16$.
The component $\Delta \tau_{\phi\phi} = \tau_{\phi\phi}-\overline{\tau_{\phi\phi}}|_{t_s}^{t_e}$ is given relative to the mean value
between times $t_s=2$ and $t_e=10$.
Snapshot of the stress field is taken at time $t=21.54$ and it displays \rev{weakly chaotic flow} (see Sec. \ref{sec:ET}). The video shows the transition from the laminar flow to the \rev{weakly chaotic flow} and ultimately elastic turbulence.
}
  \label{fig:TransitionWi12.16}
\end{figure*}

\begin{figure*}
\centering
\includegraphics[width=0.72\textwidth]{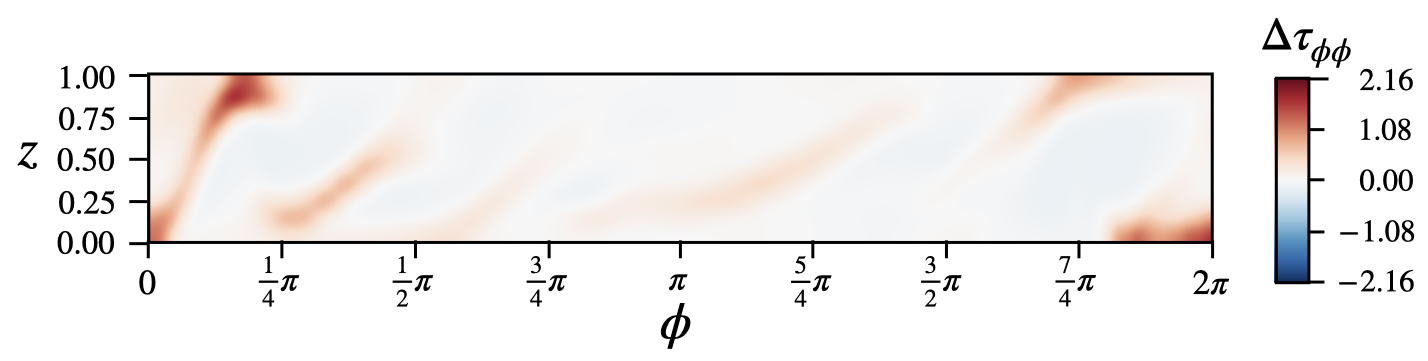}
\caption{(Multimedia view) Azimuthal stress-tensor component color-coded
in the $z-\phi$ plane near the outer radius $r=0.99\,r_o$ at $\mathrm{Wi}=12.92$.
The component $\Delta \tau_{\phi\phi} = \tau_{\phi\phi}-\overline{\tau_{\phi\phi}}|_{t_s}^{t_e}$ is given relative to the mean value between times $t_s=2$ and $t_e=3$.
Snapshot of the stress field is taken at time $t=19.63$ and it displays elastic turbulence (see Sec. \ref{sec:ET}). The video shows the transition from the laminar flow to the \rev{weakly chaotic flow} and ultimately elastic turbulence.
}
  \label{fig:TransitionWi12.92}
\end{figure*}

Likewise, we observe similar spatiotemporal behavior at $\mathrm{Wi}=12.16$ and $\mathrm{Wi}=12.92$.
We present videos of the color-coded stress component $\tau_{\phi\phi}$ in the $z-\phi$ plane near the outer cylinder ($r=0.99 \,r_o$) as a function of time in Fig.~\ref{fig:TransitionWi12.16} (Multimedia view) for $\mathrm{Wi}=12.16$ and in Fig.~\ref{fig:TransitionWi12.92} (Multimedia view) for $\mathrm{Wi}=12.92$.
The videos show the same development of the instability as the one above. An unstable nonaxisymmetric mode with four-fold symmetry drives the laminar flow towards an \rev{weakly chaotic flow} and ultimately to a chaotic flow, which we characterize as turbulent in the next section.
The snapshot presented in Fig.~\ref{fig:TransitionWi12.16} represents \rev{weakly chaotic flow}, while the snapshot presented in Fig.~\ref{fig:TransitionWi12.92} represents elastic turbulence.
Both flow states are discussed in detail in the next section.

The obvious difference between the  linear stability analysis, which predicts the $|m|=3$ {modes}
as the most unstable {modes,}
and the result of the numerical simulations can be due to simplifications used in the linear stability analysis. 
In the ansatz functions (\ref{eq:linstab1}) to (\ref{eq:linstab3}) the spatial dependence on $z$ and $r$ is decoupled and an infinitesimal flow field is assumed, whereas in the numerical solutions a more general variation in $z$, $r$ is allowed.
Additionally, the von \Karman geometry has a finite extension with an additional boundary condition for the flow field at the outer 
cylindrical boundary ($u_r = 0$ at $r=r_o$). Moreover, the spatial discretization used in the simulations is constructed with four-fold symmetry (\rev{starting from four connected trapezoidal prisms})
leading to small numerical errors, which reflect this symmetry.
This could result in lowering the critical conditions for the $|m|=4$ mode.
To analyze the influence of the mesh symmetry, we performed additional simulations with a spatial discretization with three-fold symmetry 
\rev{(starting from three connected trapezoidal prisms).}
\rev{These simulations}
at $\mathrm{Wi} = 11.78$, $\mathrm{Wi} = 12.16$ and $\mathrm{Wi} = 12.92$ all show a non-axisymmetric pattern with four-fold symmetry developing from the base flow. Therefore, we conclude that the observed difference in the most unstable mode is not due to the symmetry of the mesh discretization.

\subsection{Characterizing elastic turbulence}
\label{sec:ET}

\begin{figure*}
\centering
\includegraphics[width=.85\textwidth]{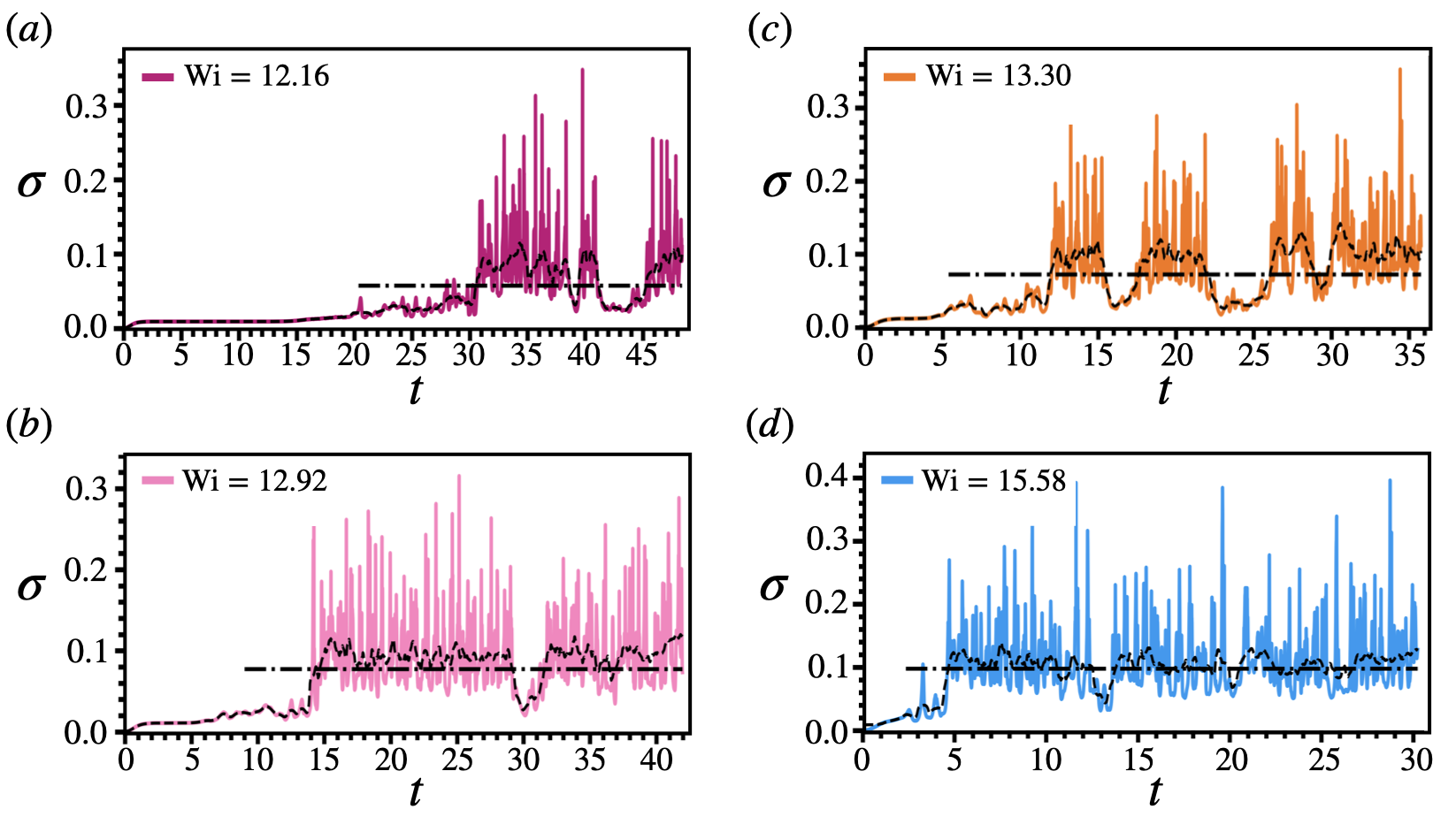}
\caption{
    Secondary-flow strength $\sigma$ defined in Eq. (\ref{eq.standard}) plotted as a function of time $t$ in units of $2\pi / \Omega$. The Weissenberg number is (a) $\mathrm{Wi}=12.16$, (b) $\mathrm{Wi}=12.92$, (c) $\mathrm{Wi}=13.3$, and (d) $\mathrm{Wi}=15.58$. The temporal average of the secondary-flow strength, $\overline\sigma(t > t_0)$, is displayed with the black dash-dotted line and starts at $t_0$, which is the first time with $\sigma \geq 0.025$. The moving average, $\sigma_\delta (t) = \frac{1}{2\delta}\int_{t-\delta t}^{t+\delta t} \sigma(t') dt' $ with $\delta t = 0.3$, is indicated by the black dashed line. 
    Two distinct flow states are distinguished: a turbulent state where $\sigma(t)$ strongly fluctuates around $\langle \sigma \rangle \approx 0.1$ and an \rev{weakly chaotic} time-dependent state with only small fluctuations of $\sigma(t)$ around a significantly smaller mean value.
    The state of the fluid flow can be identified as turbulent ($\sigma_\delta > \overline\sigma$), as \rev{weakly chaotic} ($\sigma_\delta < \overline\sigma $), or as laminar ($\sigma \approx 0$).
    }
\label{fig:SecFlow}
\end{figure*}

Here we demonstrate the occurrence of an elastic instability 
which ultimately develops into elastic turbulence. The onset of the elastic instability is indicated by the critical Weissenberg number $\mathrm{Wi_c}$.
To investigate and characterize the transition, we define an order parameter 
$\Phi = \mean{\sigma(t)}$, as the time average of the normalized velocity fluctuations 
\begin{equation}
\sigma(t) \equiv\left. \sqrt{\left\langle\left[\bm{u}(\bm{r},t) -
{ \bm{u}^0(\bm{r}) }
\right]^2\right\rangle}\right/u^0_{\mathrm{max}} \, ,
\label{eq.standard}
\end{equation}
relative to the base flow field $\bm{u}^0(\bm{r})$, 
for which we choose the numerical solution at $\mathrm{Wi}=0$. It deviates from the flow field of Eq.\ (\ref{eq:baseflow}) since the bounding parallel plates are not infinitely extended.
Here $\langle \dots \rangle = \frac{1}{V} \int \dots dV $ denotes the volume average and $u^0_{\mathrm{max}}= \Omega r_o$ is the maximum velocity of the rotating upper plate, which is used to normalize $\sigma$.
We call $\sigma(t)$ the secondary-flow strength.
Our order parameter is similar to the turbulence kinetic energy $\epsilon$, the time-averaged kinetic energy of the velocity fluctuations, which reads
\begin{equation}
\epsilon = \frac{1}{2} \overline{\left\langle\left[\bm{u}(\bm{r},t)-\overline{\bm{u}(\bm{r},t)}\right]^2\right\rangle} \, .
\end{equation}
However, both parameters scale differently; while $\epsilon \sim \bm{u}^2$, our order parameter $\Phi \sim \bm{u}$.
It is a linear measure for the relative strength of the velocity fluctuations and we use it to quantify the deviation from the base flow when a secondary flow occurs.

\subsubsection{Velocity fluctuations}
In previous work we have shown an elastically driven instability in a two-dimensional Taylor-Couette flow, where the order parameter sharply increases beyond a critical Weissenberg number $\mathrm{Wi_c}$ \cite{vanBuel2018elastic}. 
Now, we characterize this transition in the three-dimensional geometry of the von \Karman swirling flow.
In Fig.~\ref{fig:SecFlow}~(a)-(d) we plot the secondary-flow strength $\sigma$ 
as a function of time $t$ for four different Weissenberg numbers in the range $12.16 \leq \mathrm{Wi} \leq 15.58$ starting from a fluid
{at rest}
at $t=0$.
From the figure it can be observed that both the fluctuations of the secondary-flow strength and its mean, indicated by the dash-dotted line, increase with increasing \Wi.
Moreover, $\sigma(t)$ reveals bursts of strong velocity fluctuations interrupted by more quiescent flow. Thus, we distinguish two types of time-dependent flows: the first, which we denote 
{\rev{weakly chaotic}}, 
comprises small fluctuations in $\sigma$ around a small mean value, while the second turbulent branch shows large fluctuations in $\sigma$ around a mean value $\langle \sigma \rangle \approx 0.1$.

For Weissenberg numbers in the range $12 \leq \mathrm{Wi} < 20 $ the flow is bistable, it randomly switches between both flow states. 
Determining the time where steady-state is reached is difficult, since the mean value of $\sigma$ depends on how often the flow randomly switches between both states.
Therefore, in this work we have the following convention: the time-dependent 
{\rev{weakly chaotic}}
flow with its small fluctuations starts when $\sigma$ exceeds the value 0.025, which is chosen by visual inspection. The time of the first occurrence of $\sigma \geq 0.025$ is denoted $t_0$ and here we start with the averaging of $\sigma$.
An important observation in Fig.~\ref{fig:SecFlow} is that the instability takes significantly longer to set in at Weissenberg numbers closer to $\mathrm{Wi}=12$, i.e., $t_0$ decreases with increasing \Wi. For example, at $\mathrm{Wi} = 12.16$ it takes about 20 rotations (corresponding to $t_0\approx 100 \lambda$), whereas at $\mathrm{Wi} = 15.58$ it takes about 3 rotations (corresponding to $t_0\approx 12 \lambda$).

\subsubsection{Order parameter}
\label{sec:op}
\begin{figure}
\centering
\includegraphics[width=0.45\textwidth]{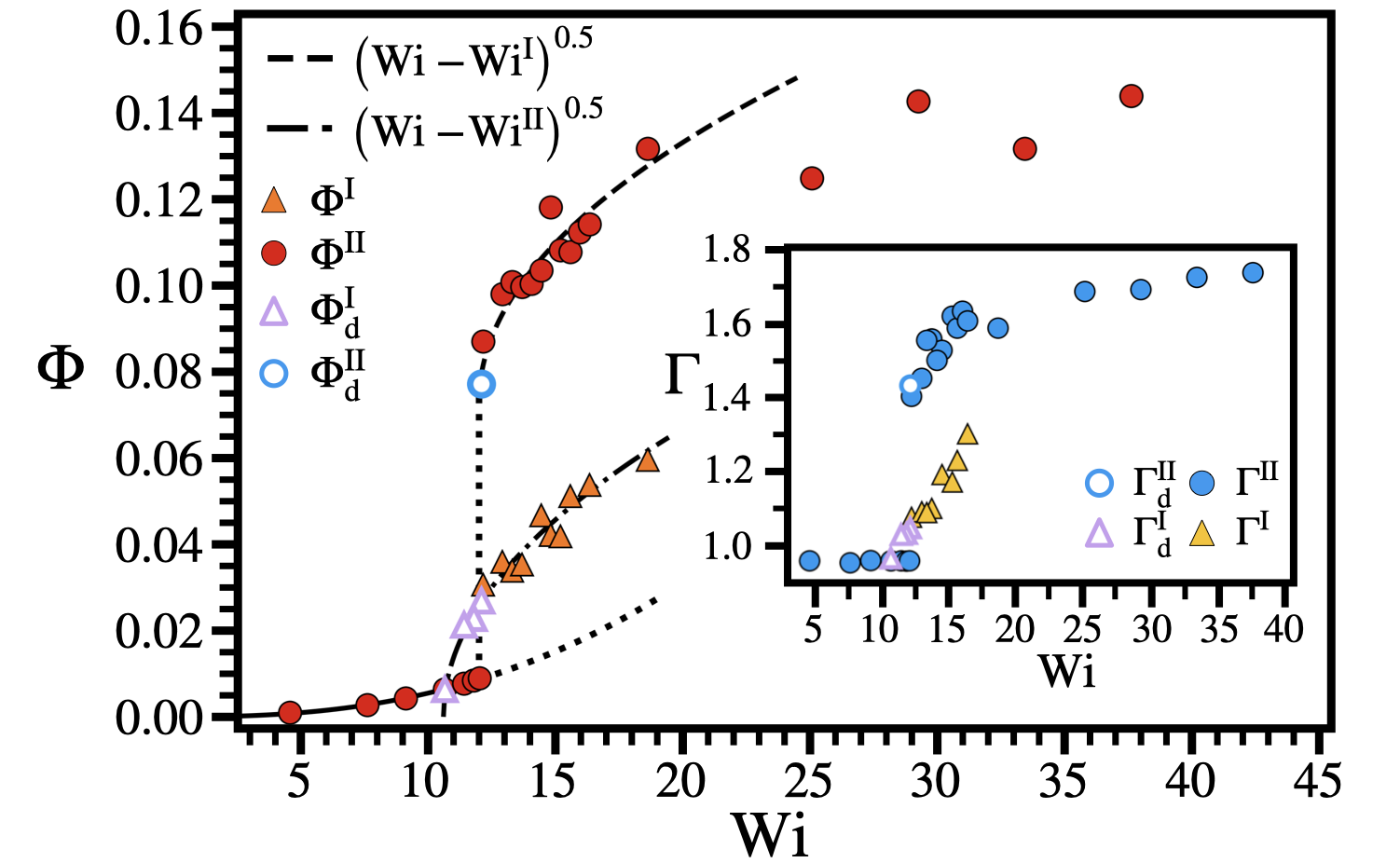}
\caption{
    Order parameter $\Phi= \mean{\sigma}$ as a function of the Weissenberg number \rev{ $\mathrm{Wi}=\Omega\lambda r_o/d$}. Closed symbols indicate the order parameter as defined in Eq.\ (\ref{eq:opdef}) for simulations starting with a fluid at rest. The red circles denote the turbulent branch $\Phi^{\mathrm{II}}$ and the orange triangles denote the \rev{weakly chaotic} branch $\Phi^{\mathrm{I}}$.  
    Below the subcritical transition at $\mathrm{Wi_c}=12$ both branches merge and 
    $\Phi^{\mathrm{I}} = \Phi^{\mathrm{II}}$.
    The dashed and dash-dotted lines show fits to square root scaling close to $\mathrm{Wi_c}$ using $\Phi^{\mathrm{I}} \sim \left({\mathrm{Wi}-\!\mathrm{Wi^\mathrm{I}}}\right)^{0.5}$ with $\mathrm{Wi}^\mathrm{I} = 10.6$ and $\Phi^{\mathrm{II}} \sim \left({\mathrm{Wi}-\!\mathrm{Wi^\mathrm{II}}}\right)^{0.5} + \Phi_0$ with $\mathrm{Wi}^\mathrm{II} = 12$ and $\Phi_0 = 0.073$, respectively.
    Dotted lines are given as guides to the eye.
    Open symbols indicate the order parameters $\Phi^\mathrm{I}_\mathrm{d}$ and $\Phi^\mathrm{II}_\mathrm{d}$ when the simulations run for 26 rotations at $\mathrm{Wi}=15.96$ and then \Wi is dropped to its actual value. Hysteretic behavior is observed for the lower branch.
    Inset: Flow resistance $\Gamma$ for the turbulent (II) and \rev{weakly chaotic} (I) branch. }
  \label{fig:orderparameter}
\end{figure}

To capture and describe the two different time-dependent flow states, we cannot simply look at  the order parameter, $\Phi = \mean{\sigma}$, as the total time average over $\sigma$.
Moreover, unless the total simulation time approaches infinity, $\Phi$ strongly depends on the time where averaging starts.
Therefore, we need a more careful treatment. First, we properly define the mean secondary-flow strength, ${\sigma_0} = \mean{\sigma(t > t_0)}$, by starting the average at $t_0$. In Fig.~\ref{fig:SecFlow} it is indicated as the black dash-dotted line starting at $t=t_0$.
Additionally, a moving average $\sigma_\delta$ is displayed in Fig.~\ref{fig:SecFlow} by the black dashed line.
The moving average is defined as $\sigma_\delta (t) = \frac{1}{2\delta}\int_{t-\delta t}^{t+\delta t} \sigma(t') dt' $, where we have set $\delta t = 0.3$.
\rev{At the start and end of the time series the integral is limited to the range of $\sigma(t)$.
\rev{Thus,}
the normalization factor $(2\delta)^{-1}$ is adjusted to the range of the integral, accordingly.}
Now, we are able to introduce order parameters for the two flow states of the fluid by averaging only over time traces $\sigma(t)$, which belong to either the turbulent (II) or \rev{weakly chaotic} (I) flow state:
\begin{equation}
\Phi=
\begin{cases}
\Phi^\mathrm{II} = \mean{\sigma(t)}  & 
\enspace \text{with} \enspace \sigma_\delta(t) \geq {\sigma_0}, \enspace  t > t_0,
\\
\Phi^\mathrm{I\,} = \mean{\sigma(t)}  & 
\enspace \text{with} \enspace  \sigma_\delta(t) < {\sigma_0}, \enspace  t > t_0.
\end{cases}
\label{eq:opdef}
\end{equation}
The order parameters are plotted in Fig.~\ref{fig:orderparameter} as a function of the Weissenberg number.
A clear transition from the laminar base flow where $\Phi\approx0$ to the occurrence of two types of secondary-flows with $\Phi^\mathrm{I},\Phi^\mathrm{II}>0$ is observed upon increasing the elasticity of the fluid beyond a critical value of $\mathrm{Wi}_c = 12$.
Moreover, the order parameters for both flow types show a subcritical transition upon increasing the Weissenberg number beyond the critical value. A subcritical transition is also predicted in theory for the \mbox{Oldroyd-B} model by linear stability analysis \cite{walsh1987flow,oztekin1993instability}
and observed in experiments \cite{groisman2004elastic,burghelea2007elastic}.
Remarkably, our observed $\mathrm{Wi}_c$ is close to the critical value $\mathrm{Wi}=12.5$ observed in experiments with the 
same aspect ratio $d/r_o$ but at a higher polymer viscosity $\beta = 0.8$ \cite{burghelea2007elastic}.

\subsubsection{Flow resistance}

Another characteristic of elastic turbulence is a sharp increase in the flow resistance 
\cite{groisman2000,groisman2004elastic,vanBuel2018elastic}.
We define the flow resistance using the total work performed on the fluid per unit time and unit volume,
\begin{equation}
    W = \frac{1}{V} \oint \bm{u} \cdot \bm{T} \bm{n} \, dA \, ,
\end{equation}
where $\bm T = -p \bm I +  \eta_s [\nabla \otimes \bm{v} + (\nabla \otimes \bm{v})^\mathrm{T}] + \bm\tau$ is the Cauchy stress tensor, $\bm{n}$ is the surface normal vector, $V$ is the volume, and we integrate over the total surface $A$ of the fluid.
$W$ has both a viscous and elastic contribution, $W = W_\mathrm{v} + W_\mathrm{el}$. For an elastically driven instability we expect 
$W_\mathrm{el} > W_\mathrm{v}$, since we expect the elastic 
{stresses} 
to sharply increase after the transition compared to the velocity gradients in the flow.
The elastic contribution to the total work results from the polymeric stress tensor and
for the von \Karman swirling flow reads
\begin{equation}
    W_\mathrm{el} = \frac{1}{V} \int u_\phi \tau_{z\phi} dA_t + \frac{1}{V} \int (u_\phi \tau_{r\phi} + u_z \tau_{rz} ) dA_s \, ,
\end{equation}
where $A_t$ is the area of the fluid at the upper plate \mbox{($\bm{n} = \bm{e}_z$)} and $A_s$ is the area of the fluid at the side {($\bm{n} = \bm{e}_r$).}
Note that our boundary condition sets $u_r = 0$ at the side and at the top plate $u_\phi = r \Omega$ is the only non-zero velocity component.
Likewise, the viscous contribution from the solvent is given by
\begin{equation} 
W_\mathrm{v} = \frac{2\eta_s}{V} \int u_\phi D_{z\phi} dA_t 
+ \frac{2\eta_s}{V} \int (u_\phi D_{r\phi} + u_z D_{rz} ) dA_s \, ,  
\end{equation}
where $\bm D = \frac{1}{2}[\nabla \otimes \bm{u} + (\nabla \otimes \bm{u})^\mathrm{T}]$ is the deformation rate tensor.
For the laminar base flow, excluding edge effects, we have 
\begin{align}
W_\mathrm{el}^0= \frac{\eta_p}{2d}\left(\frac{\Omega r_o}{d}\right)^2
\quad
\rev{\text{and}}
\quad \enspace
W_\mathrm{v}^0
= \frac{\eta_s}{2d}\left(\frac{\Omega r_o}{d}\right)^2 \, .    
\label{eq.work_reference}
\end{align}

Now we define the flow resistance $\Gamma = \overline{W(t)} / W^0$, where the temporal mean work $\overline{W(t)}$ is normalized by the laminar-base-flow value $W^0$.
Corresponding to the two time-dependent fluctuating flow states, we also introduce the flow resistances of the turbulent (II) and \rev{weakly chaotic} (I) state:
\begin{align}
\Gamma=
\begin{cases}
\Gamma^\mathrm{II} = \mean{W(t)}/W^0  &
\enspace \text{with} \enspace \sigma_\delta(t) \geq {\sigma_0}, \! \enspace  t > t_0,
\\
\Gamma^\mathrm{I}\; = \mean{W(t)}/W^0  & 
\enspace \text{with} \enspace  \sigma_\delta(t) < {\sigma_0}, \! \enspace  t > t_0.
\end{cases}
\label{eq:FRdef}
\end{align}
We plot $\Gamma$ in the inset of Fig.~\ref{fig:orderparameter}.
A clear and sudden increase in the flow resistance beyond the critical Weissenberg number can be observed, associated with the subcritical transition indicated be the order parameter.
At the transition to the turbulent branch (filled blue circles above $\mathrm{Wi_c}$) the flow resistance jumps by $0.4\, \Gamma_0$, which is below the experimentally observed value \cite{burghelea2007elastic}. For the \rev{weakly chaotic flow}
(filled {yellow} triangles) the jump is $0.1\, \Gamma_0$, four times less than the turbulent branch, and the normalized flow resistance goes up to 1.3.
For \mbox{$\mathrm{Wi} > 18$} the \rev{weakly chaotic flow} state is no longer
observed and therefore only the turbulent branch remains.
The total value of the flow resistance $\Gamma$ as a function of \Wi is in the range $1 < \Gamma < 2$, which is about half of the values observed in experiments \cite{burghelea2007elastic}.
For \mbox{$\mathrm{Wi} > 18$} the flow resistance $\Gamma$ becomes nearly constant
in our simulations. This might be due to discretization errors such that simulations at higher \Wi require finer meshes. Accurately resolving this discrepancy is beyond our current capabilities.

\begin{figure*}
  \rev{
\centering
\includegraphics[width=0.85\textwidth]{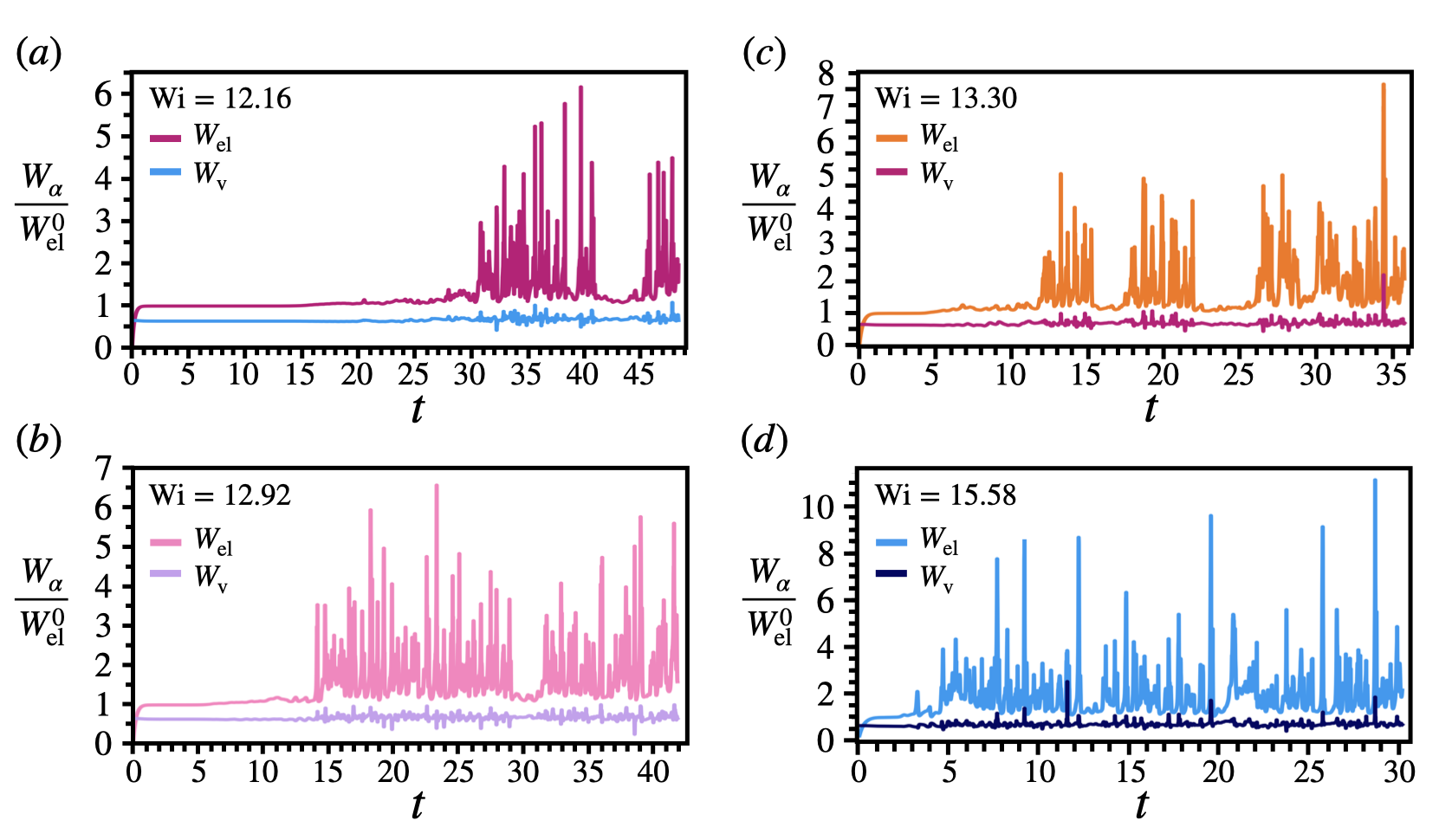}
  \caption{
  Work $W_\alpha$ performed on the viscoelastic fluid by the viscous stresses $W_\mathrm{v}$ and the elastic stresses $W_\mathrm{el}$ plotted versus time and normalized by the base flow value $W^0_\mathrm{el}$ 
 [see Eqs.\ (\ref{eq.work_reference}), left].
  The index $\alpha$ refers to the elastic ($W_\mathrm{el}$) or viscous ($W_\mathrm{v}$) contribution, respectively.
  Data are shown for four values of the Weissenberg number with (a) $\mathrm{Wi}= 12.16$, (b) $\mathrm{Wi}= 12.92$, (c) $\mathrm{Wi}= 13.30$ and (d) $\mathrm{Wi}= 15.58$.
  }
  \label{fig:FR}
  }
\end{figure*}

\begin{figure}
\centering
\includegraphics[width=0.39\textwidth]{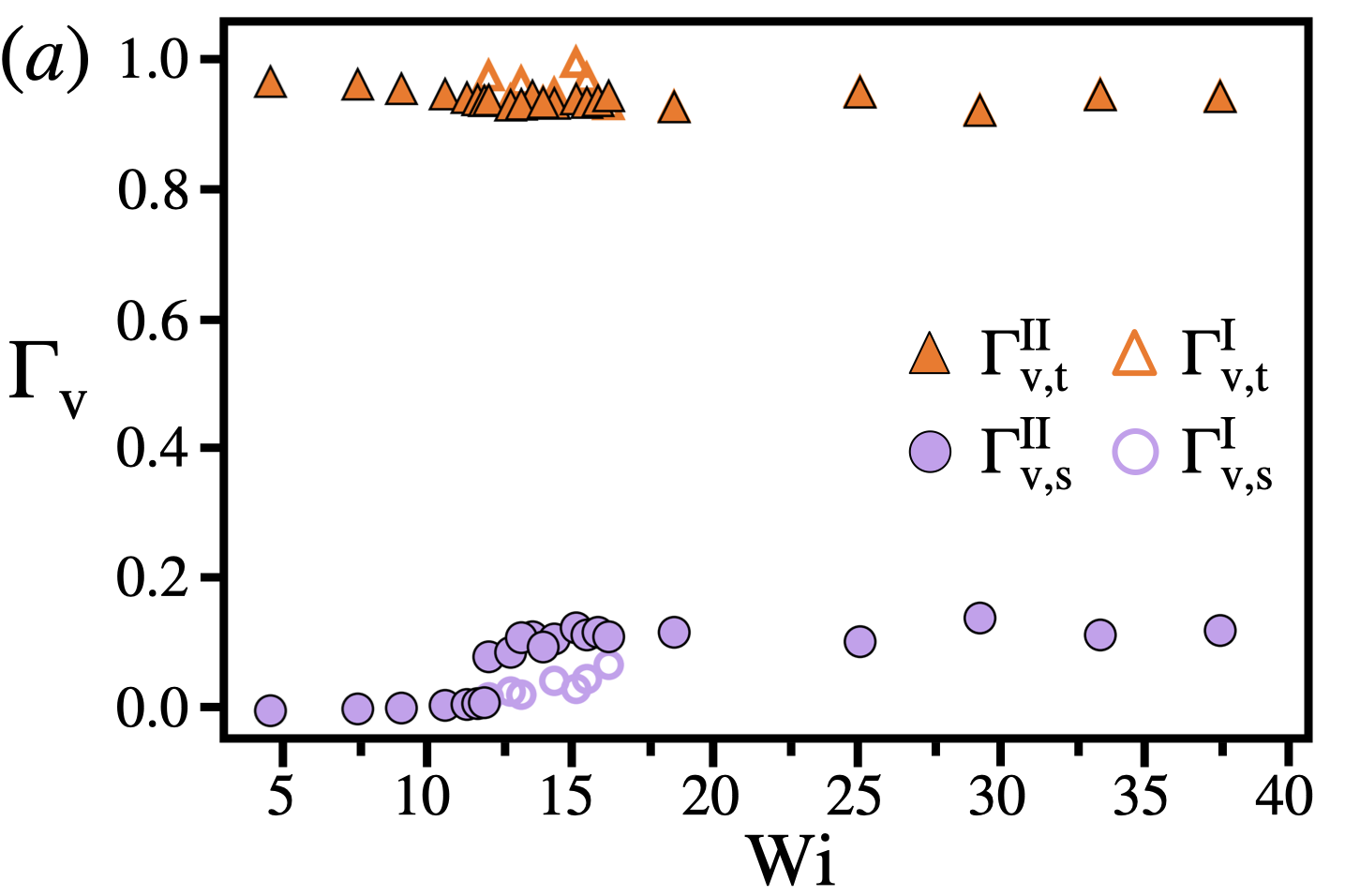}
\includegraphics[width=0.39\textwidth]{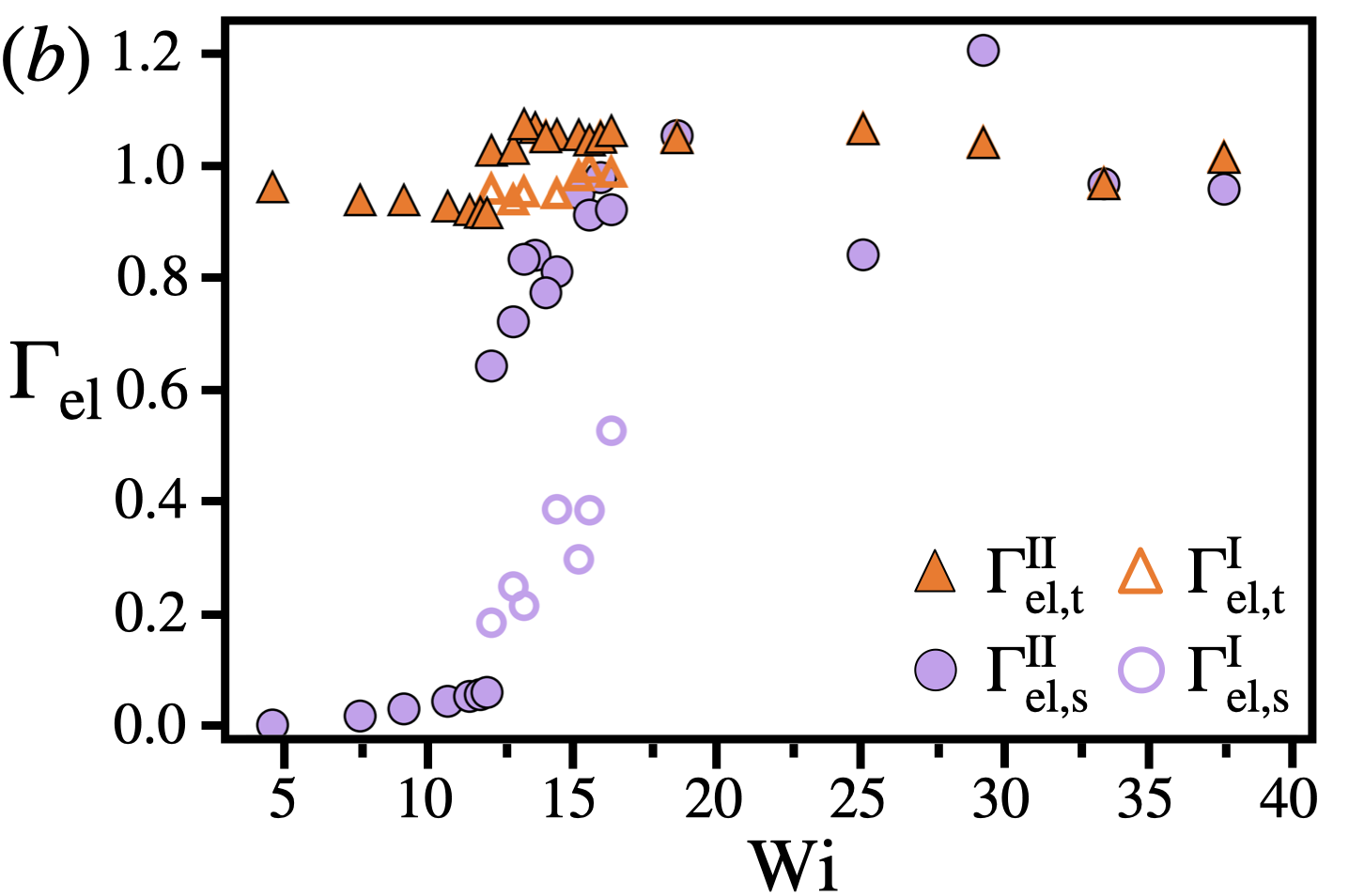}
  \caption{
  { (a) Viscous part of the flow resistance $\Gamma_{\mathrm{v},\nu}^{\mathrm{I/II}} = \mean{W_{\mathrm{v},\nu}} / W_\mathrm{v}^0$  
  and (b) elastic part of the flow resistance $\Gamma_{\mathrm{el},\nu}^{\mathrm{I/II}} = \mean{W_{\mathrm{el},\nu}} / \rev{W_\mathrm{el}^0}$   
  plotted versus \Wi. The subscripts $\nu =t,s$ refer to the top (orange triangles) and side surface (purple circles), respectively. 
  The superscripts I, II refer to \rev{weakly chaotic} (open symbols) and turbulent/laminar flow (closed symbols), respectively.}
}
  \label{fig:Gamma}
\end{figure}

\subsubsection{Elastic and viscous work}

Next, we discuss the total work $W$ in more detail. We first analyze the viscous ($W_\mathrm{v}$) and elastic ($W_\mathrm{el}$) parts relative to the base flow values and then evaluate the contributions from the two surfaces $A_t$ and $A_s$.

\rev{The normalized viscous and elastic contributions to the total work per unit time, 
$W_\mathrm{v}/W_\mathrm{el}^0$ and $W_\mathrm{el}/W_\mathrm{el}^0$, are plotted versus time in Fig.~\ref{fig:FR} for different $\mathrm{Wi}$.
As reference we have chosen the base flow value $W_\mathrm{el}^0$ from Eq.\ (\ref{eq.work_reference}), which is related to the viscous base flow value $W_\mathrm{v}^0$
by $W_\mathrm{el}^0 = \beta \, W_\mathrm{v}^0$.
At $\mathrm{Wi}=12.16$ [plot (a)], just above the critical Weissenberg number, 
we only observe slight fluctuations in the weakly chaotic state,
whereas large fluctuations are present in the turbulent state.
This is in full analogy to the secondary flow strength, presented in Fig.\ \ref{fig:SecFlow}.
The fluctuations in the turbulent state of both the viscous and the elastic stresses increase in magnitude for higher \Wi.
Before the onset of the elastic instability, in the laminar flow state, the viscous work  
$W_\mathrm{v}/W^0_\mathrm{el}  = 1/\beta = 2/3$ and the elastic work $W_\mathrm{el}/W^0_\mathrm{el} = 1$ are equal to their respective base flow values.
}
Generally, the magnitude of the fluctuations in $W_\mathrm{v}$ are small compared to the fluctuations in $W_\mathrm{el}$. 
The time-averaged value of $W_\mathrm{v}$ remains close to 
\rev{$\beta^{-1}$, \textit{i.e.},}
$\overline{W_\mathrm{v}}/W_\mathrm{v}^0 \approx 1$, while the average value of $W_\mathrm{el}$ increases.
Thus, $W_\mathrm{el} > W_\mathrm{v}$ as one expects for an elastically driven instability and as it is also demonstrated by Fig.~\ref{fig:FR}. Therefore, we conclude that the increase in the flow resistance $\Gamma$, the normalized total work per unit time, is mainly due to the elastic work performed on the fluid.

Next, we discuss the contributions from the different surface areas of the cylindrical geometry
to the flow resistance. We distinguish between the viscous and elastic contributions of the work performed on the top and the side surface areas.
We do not need to consider the bottom area since the work is always zero ($\bm{u} = 0$).
The two unsteady flow states, turbulent (II) and \rev{weakly chaotic} (I), combined with the two surface areas give four specific contributions to the flow resistance, which we describe by
\begin{equation}
\Gamma_{\alpha,\nu} =
\begin{cases}
\Gamma^\mathrm{II}_{\alpha,\nu} = \mean{W_{\alpha,\nu}(t)}/W_{\alpha}^0 \enspace 
\enspace \text{with} \enspace \sigma_\delta(t) \geq {\sigma_0}, \enspace  t > t_0,
\\
\Gamma^\mathrm{I}_{\alpha,\nu} = \mean{W_{\alpha,\nu}(t)}/W_\alpha^0 \enspace 
\enspace \text{with} \enspace  \sigma_\delta(t) < {\sigma_0}, \enspace  t > t_0,
\end{cases}
\label{eq:FRelvisc}
\end{equation}
where in the following, we will use {the} superscript II also for the laminar state.
Here, $\alpha$ refers to the elastic ($\Gamma_\mathrm{el}$) or viscous ($\Gamma_\mathrm{v}$) contribution,
while the index $\nu$ labels the the side surface $A_s$ with $\nu=\mathrm{s}$ and the top surface $A_t$
with $\nu=\mathrm{t}$. The work performed on the side surface in the laminar state is small, as we show below. In fact for the analytic expression of Eq.\ (\ref{eq:baseflow}) it is zero.

In Figs.\ \ref{fig:Gamma}(a) and (b) the flow resistances for the viscous $\Gamma^\mathrm{I/II}_{\mathrm{v},\nu}$ and elastic $\Gamma^\mathrm{I/II}_{\mathrm{el},\nu}$ part are plotted versus \Wi, respectively. Closed symbols refer to turbulent/laminar flow (II) and open symbols to the \rev{weakly chaotic flow} state (I). Orange triangles indicate the contribution from the top plate ($\nu=\mathrm{t}$) and purple circles from the
side surface {($\nu=\mathrm{s}$)}.
Figure\ \ref{fig:Gamma}(a) shows that the viscous flow resistance in all flow states is dominated by the contribution from the top plate
(triangles) with a nearly constant value $\Gamma^\mathrm{I/II}_{\mathrm{v},t} \approx 1$, while the side surface contributes 
at most around 12 \%.
The viscous flow resistance from the side surface jumps from a nearly zero value in the laminar state to around $0.12\,W_\mathrm{v}^0$ in the turbulent 
state (closed symbols)
and remains constant for \Wi$\geq 15$, while the 
flow resistance 
{of the \rev{weakly chaotic flow}}
(open symbols) slowly increases to the same maximum value.
{The elastic flow resistance plotted in Fig.~\ref{fig:Gamma}(b) shows similar behavior.
The contribution from the top plate roughly stays at $\Gamma^\mathrm{I/II}_{\mathrm{el},t} \approx 1$ (triangles). 
A closer inspection shows a slight decrease before the transition and a jump to $1.06$ at the transition, while the \rev{weakly chaotic flow} state has a slightly smaller resistance value. Very pronounced is the sharp increase of the flow resistance from the side surface (circles).
At the transition from the laminar state,
with its very small resistance, a jump to $0.6$ in the turbulent state (closed circles)
is observed.
The flow resistance then further increases with \Wi to a maximum value of about $1.06$. 
The resistance of the \rev{weakly chaotic flow} state slowly increases 
beyond $\mathrm{Wi}_c$ until it is no longer observable at \Wi$>18$.}

In summary, we 
conclude that the sharp increase of the total flow resistance $\Gamma$ beyond $\mathrm{Wi_c}$ 
(inset of Fig. \ref{fig:orderparameter}) is mainly due to the sharp increase of the elastic contribution of the work performed on the open side surface of the swirling fluid.
Furthermore, for \Wi$>18$ the contributions to the elastic flow resistance from the top and side surfaces are roughly equal and, therefore, the total flow resistance remains nearly constant (see the inset of Fig. \ref{fig:orderparameter}). 

\rev{
\begin{figure}
\centering
\includegraphics[width=0.475\textwidth]{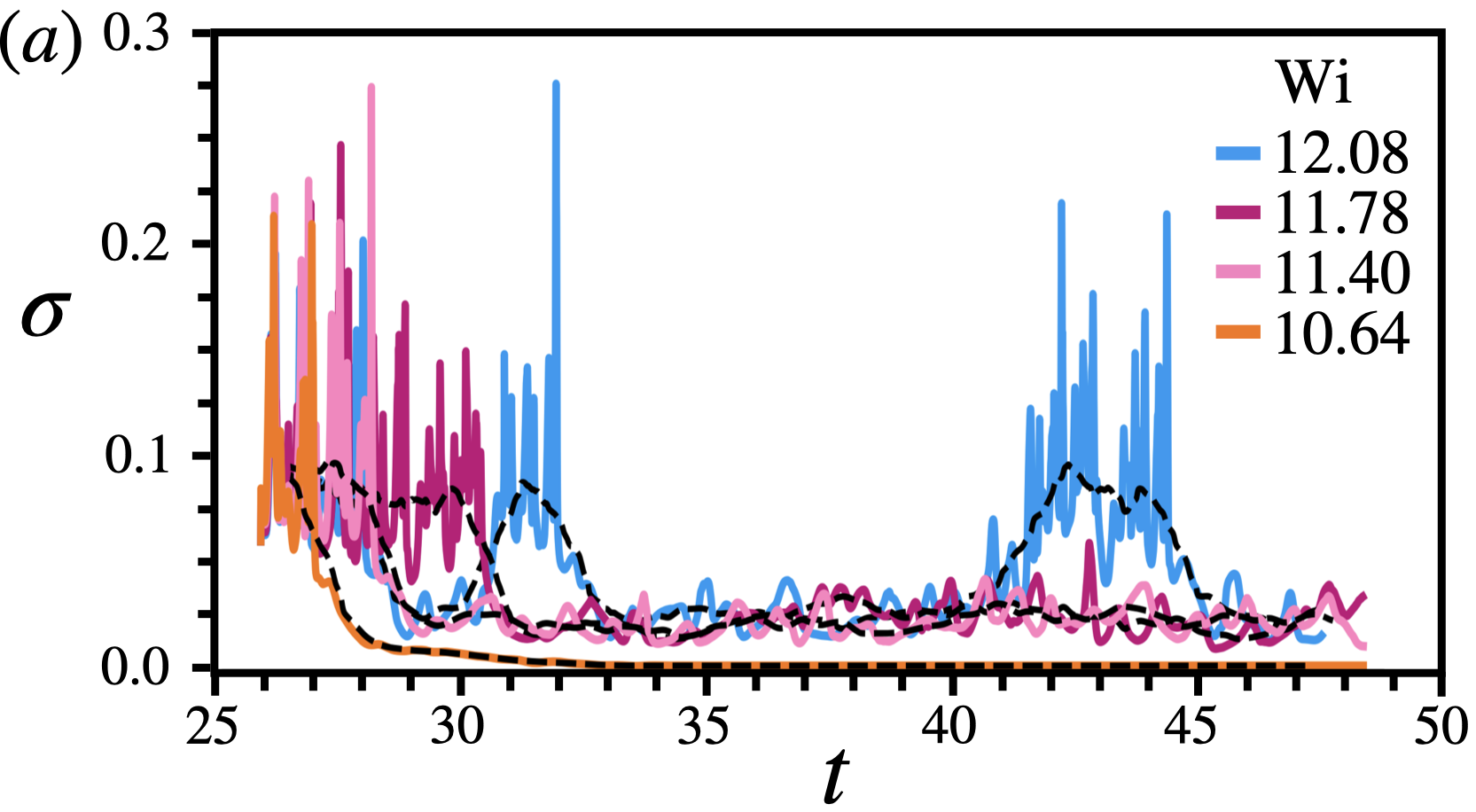}
\includegraphics[width=0.475\textwidth]{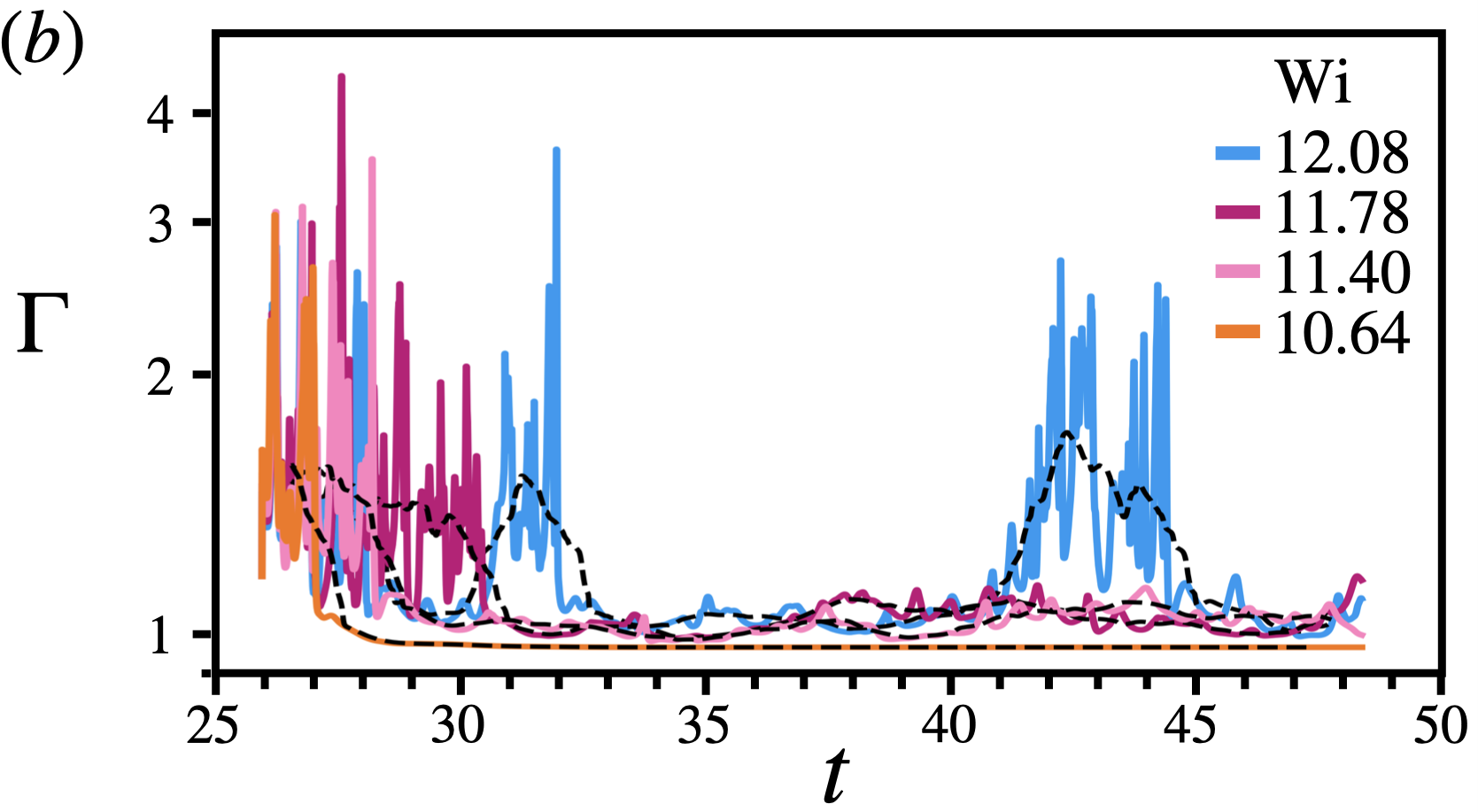}
    \caption{(a) Secondary-flow strength $\sigma$ and (b) flow resistance $\Gamma$ \rev{(log scale)} as a function of time $t$ for different \Wi when the flow is initialized in the turbulent state. In both cases the fluid flow is simulated at \mbox{\Wi = 15.96} for the first 26 rotations and then \Wi is lowered to the indicated values.
}
  \label{fig:secflowdown}
\end{figure}
}

\subsubsection{Hysteretic behavior}
\label{sec:hysteretic}

Subcritical transitions can show hysteretic behavior. To explore this further, we perform simulations which start in the turbulent flow state at a high Weissenberg number and subsequently lower its value.
The simulations are initialized at \Wi $= 15.96$ for $t = 26$ rotations. Afterwards the Weissenberg number is lowered to $\mathrm{Wi} = 12.084, 11.78, 11.4$ and $10.64$ for the next $t = 22$ rotations. The 
resulting secondary-flow strength is plotted in Fig.~\ref{fig:secflowdown}~(a) as a function of time.
We observe three different responses: bistable flow, hysteretic 
behavior, and laminar flow.
The bistable regime is observed at \Wi = 12.084. 
The secondary-flow strength $\sigma$ initially remains in the turbulent state for 3 rotations. 
Then, the flow becomes \rev{weakly chaotic} for less than 2 rotations and switches back to the turbulent branch for about 2 rotations and back to the 
\rev{weakly chaotic flow} again. After a more quiescent episode of about 10 rotations the turbulent state is entered and so on.
The order parameters of the two flow states, $\Phi_\mathrm{d}^\mathrm{I}$ and $\Phi_\mathrm{d}^\mathrm{II}$, are displayed by the open circles and 
triangles in Fig.\ \ref{fig:orderparameter}, respectively.

At lower Weissenberg numbers, \mbox{\Wi = 11.78} and \mbox{\Wi = 11.4}, below $\mathrm{Wi}_c$ clear hysteretic behavior is observed.
The secondary-flow strength drops from the initial turbulent state to the \rev{weakly chaotic flow} state and stays there for the duration of the simulation (see also the order parameter $\Phi_\mathrm{d}$ given by the open triangles in Fig.\ \ref{fig:orderparameter}).
In contrast, in simulations starting from a resting fluid the flow remains laminar with a small secondary-flow strength of about $\sigma = 0.025$ (see Sec. \ref{sec:op} and Fig. \ref{fig:orderparameter}).
Finally, at the lowest Weissenberg number, \mbox{\Wi = 10.64}, the flow becomes laminar with a secondary-flow strength $\sigma \ll 1$.

Moreover, we also study the hysteretic behavior in the flow resistance $\Gamma$, the total work performed on the 
{viscoelastic}
fluid per unit time and unit volume, which is plotted in Fig.~\ref{fig:secflowdown}(b).
Similar to the secondary-flow strength, the same three responses are observed. In the bistable flow state, the strongly fluctuating $\sigma$ and $\Gamma$
are clearly correlated, while for the \rev{weakly chaotic flow} both quantities are small.
Finally, in the laminar state, $\sigma \ll 1$, $\Gamma \leq 1$ and both quantities are constant in time.
Note that experiments performed with the von \Karman flow geometry also demonstrate hysteretic behavior \cite{groisman2000,groisman2004elastic,burghelea2007elastic,mckinley1991observations}. 
The total work performed on the fluid shows hysteresis in the range $7 < $\Wi$< 15$ \cite{burghelea2007elastic}.
However, above \mbox{\Wi$ = 18$}, square-root scaling of the flow resistance is observed in these experiments, which we do not capture in our simulations where the flow resistance \rev{only shows a small increase.}

\begin{figure}
\rev{
\centering
\includegraphics[width=0.48\textwidth]{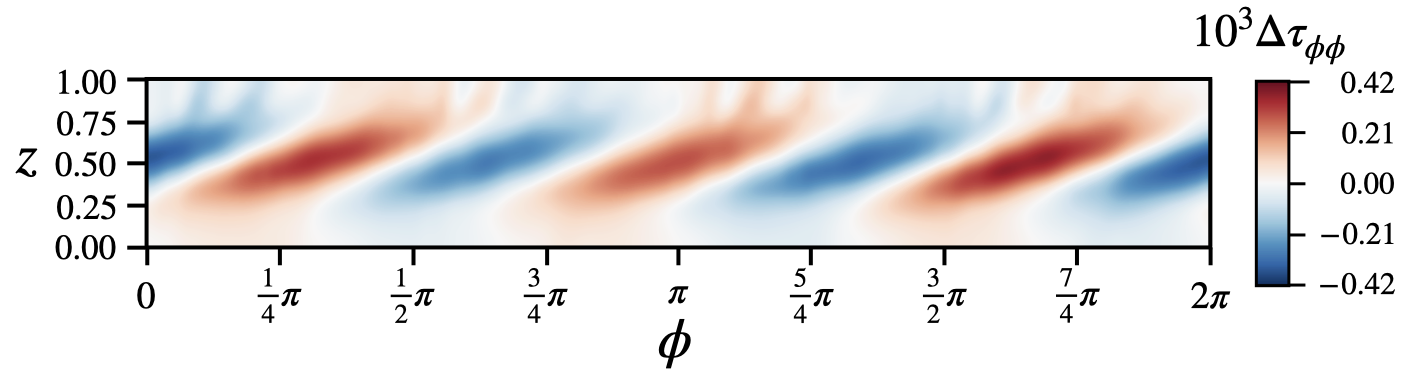}
\caption{
    (Multimedia view) Azimuthal stress-tensor component color-coded in the $z-\phi$ plane near the outer radius $r=0.99\,r_o$ at $\mathrm{Wi}=10.64$. The component $\Delta \tau_{\phi\phi} = \tau_{\phi\phi}-\overline{\tau_{\phi\phi}}|_{t_s}^{t_e}$ is given relative to the mean value calculated between times $t_s=46$ and $t_e=48$. The snapshot of the stress field is taken at time $t = 37.16$.
}
  \label{fig:TransitionDown10}
  }
\end{figure}

\rev{
We add an interesting observation about the spatial and temporal evolution of the elastic stress for the case $\mathrm{Wi}=10.64$, where the flow field develops from the full turbulent to the laminar base state.
In the turbulent state, initialized at $\mathrm{Wi}=15.96$, many spatial modes are excited, which after lowering \Wi will decay. We expect the least stable mode to decay the slowest and hence be the longest present, before the base flow is reached.
We present the spatial and temporal evolution of the color-coded elastic stress component $\tau_{\phi\phi}$ in the $z-\phi$ plane near the outer cylinder ($r=0.99 \,r_o$) in the video attached to Fig. \ref{fig:TransitionDown10} (Multi media view) for $\mathrm{Wi}=10.64$. 
The video comprises the time range, where Fig.\ \ref{fig:secflowdown}(a) indicates that the disturbance flow has vanished.
Interestingly, we still observe a weak disturbance flow corresponding to
a non-axisymmetric mode with wave number $m=\pm 3$ as the least stable mode.
In contrast, when starting the simulations with the fluid at rest, the system settles directly into the base flow.
The observed mode agrees with our linear stability analysis (see Sec. \ref{sec:lsa}), where we predicted $|m| = 3$ as the most 
unstable mode. Even the predicted critical Weissenberg number fits to the numerical results.
The laminar state is recovered at $\mathrm{Wi}=10.64< \mathrm{Wi_c^{LSA}} = 11.255 $ and the flow state remains weakly chaotic at 
$\mathrm{Wi}=11.40 >\mathrm{Wi_c^{LSA}} = 11.255 $.
}

\rev{Now, we study the spatio-temporal evolution of the elastic stress component $\tau_{\phi\phi}$ for $\mathrm{Wi}=11.40$,
where the weakly chaotic flow regime occurs. It is shown in the video attached to Fig.~ \ref{fig:TransitionDown10}(b) (Multi media view). 
Beyond $t = 29.4$ the video shows, on average, a dynamic pattern with 
three-fold symmetry. This nicely supports our linear-stability analysis and the prediction of the unstable $|m|=3$ mode beyond 
$\mathrm{Wi_c^{LSA}} = 11.255$.}

\begin{figure}
\rev{
\centering
\includegraphics[width=0.48\textwidth]{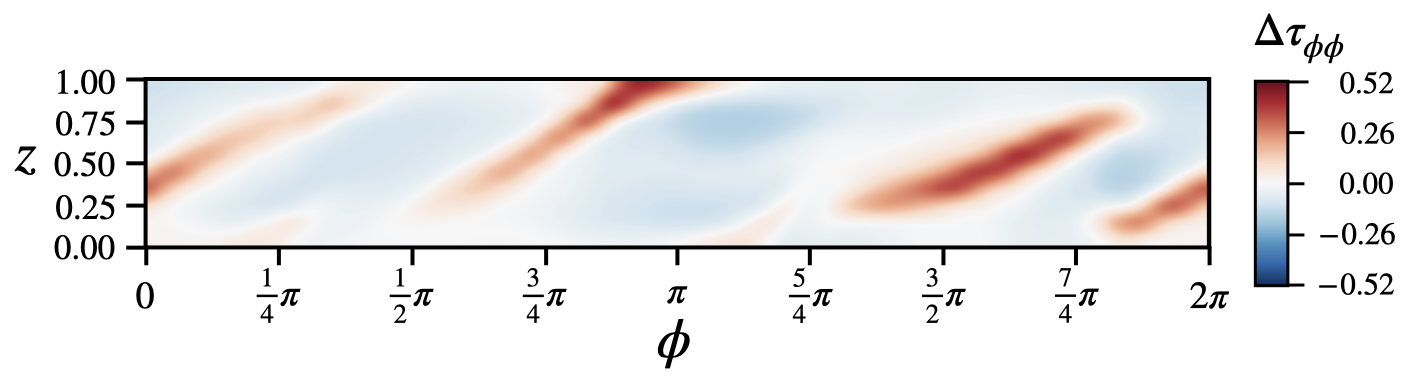}
\caption{
\rev{
    (Multimedia view) Azimuthal stress-tensor component color-coded in the $z-\phi$ plane near the outer radius $r=0.99\,r_o$ at $\mathrm{Wi}=11.40$. The component $\Delta \tau_{\phi\phi} = \tau_{\phi\phi}-\overline{\tau_{\phi\phi}}|_{t_s}^{t_e}$ is given relative to the mean value calculated between times $t_s=46$ and $t_e=48$. The snapshot of the stress field is taken at time $t = 29.4$.
    }
}
  \label{fig:TransitionDown11}
  }
\end{figure}

\subsection{Power spectral density}
\label{sec:PSD}

An important feature of elastic turbulence is the characteristic power-law dependence of the spatial and temporal power spectral 
densities of the velocity fluctuations.
We denote them by $\mathcal{P}_s (m)$ and $\mathcal{P}_t (f)$, respectively. In this section, we first discuss the temporal power spectra and secondly the spatial power spectra of the velocity fluctuations.

\subsubsection{Temporal}

\begin{figure}
\centering
\includegraphics[width=0.48\textwidth]{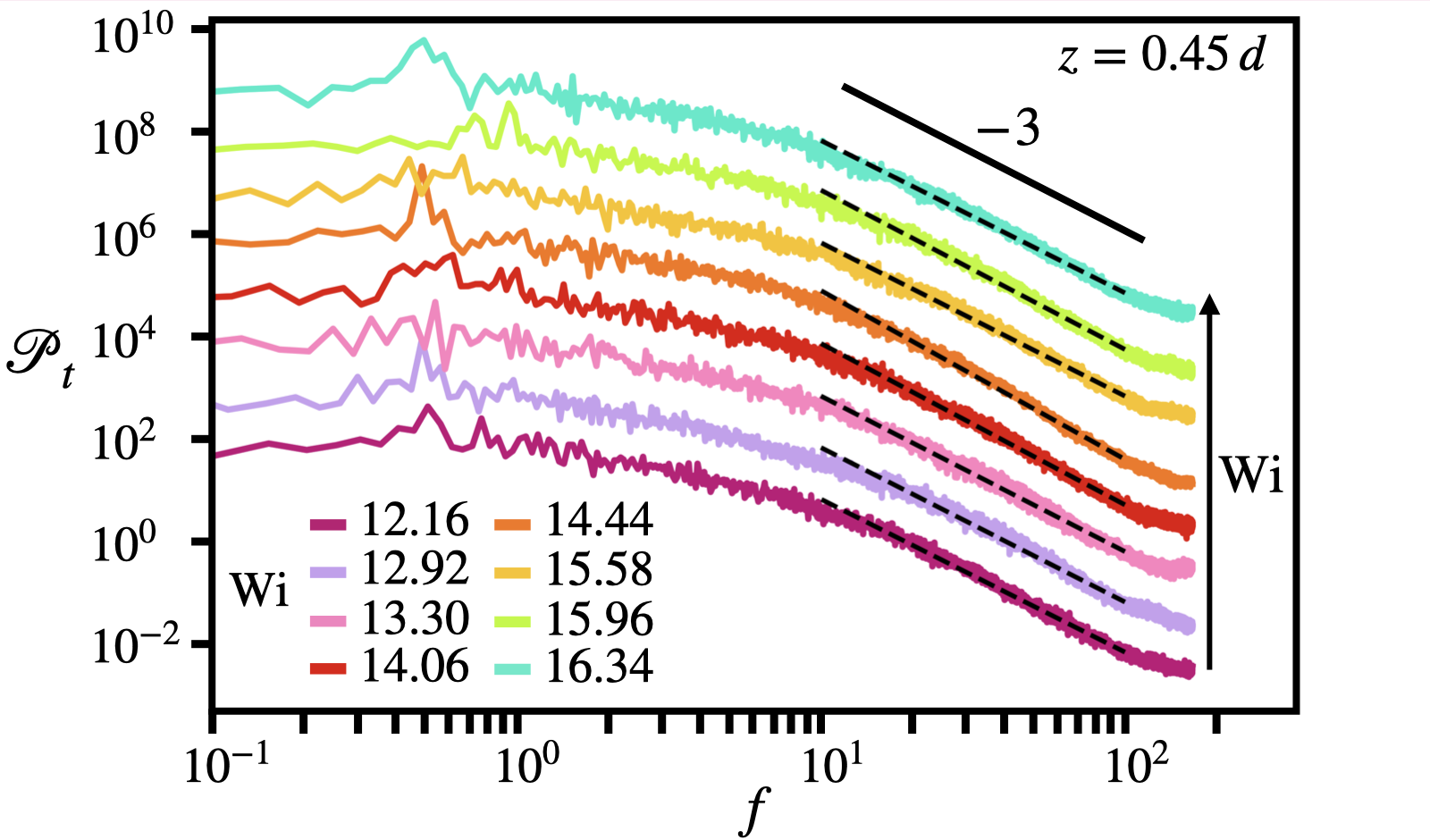}
  \caption{
  Temporal power spectral density $\mathcal{P}_t (f) = \langle | \mathcal{F}\{\bm{u}\}(f)|^2\rangle_\phi$ of the velocity fluctuations as a function of the frequency $f$ for different \Wi. The spectra are taken at height $z=0.45\, d$ and radial position $r=0.99\,r_o$. The data are arbitrarily offset for clarity by multiplying with a constant. Power laws $f^{-\alpha_t}$ are fitted and plotted as black dashed lines in the frequency range $ 9.92 \leq f  \leq 99.2$.
}
\label{fig:PSDt}
\end{figure}

\begin{figure}
\centering
\includegraphics[width=0.41\textwidth]{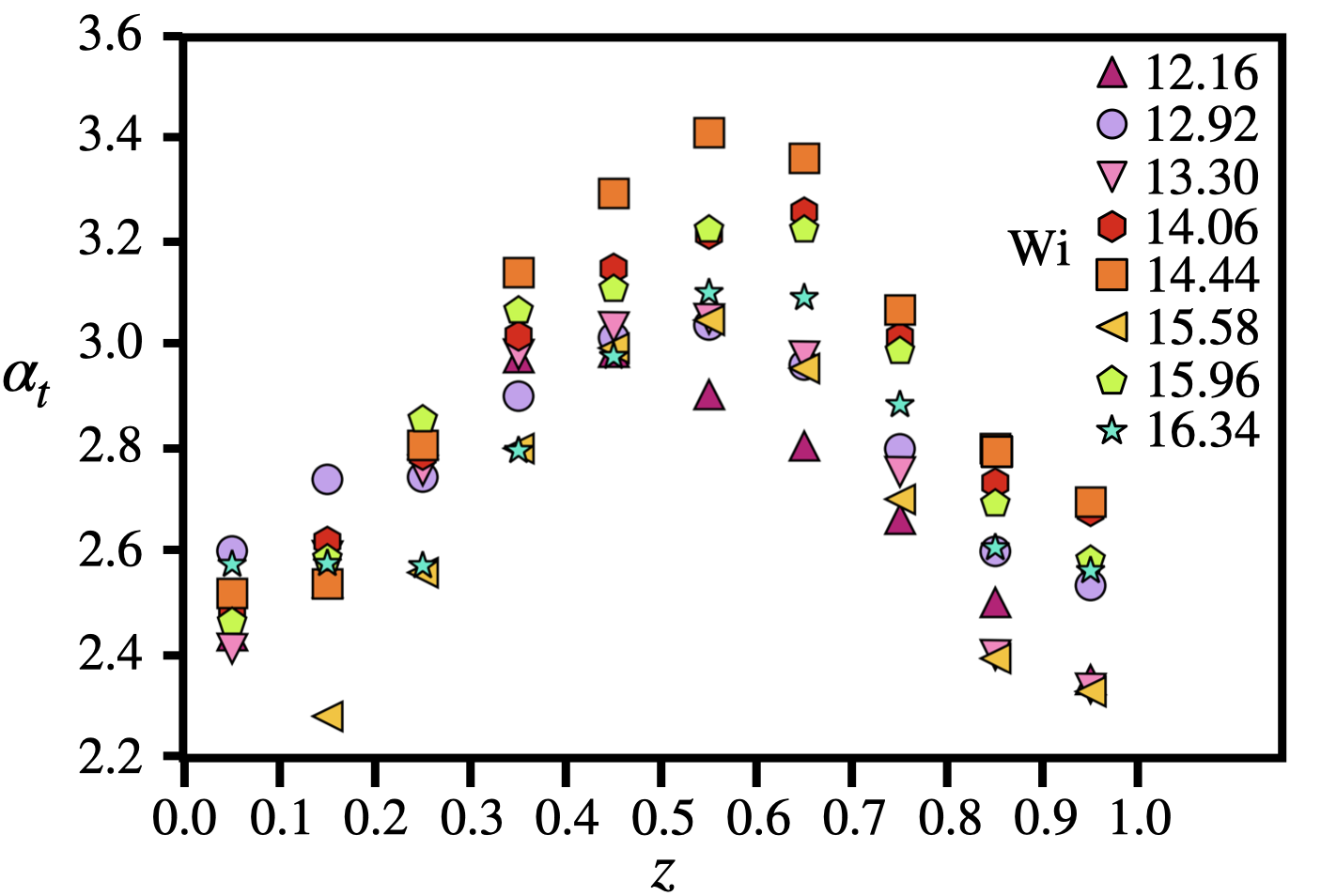}
  \caption{ Temporal exponent $\alpha_t$ plotted versus the
  height $z$ for several values of the Weissenberg number $\mathrm{Wi}$ above 
 \mbox{ $\mathrm{Wi}_c = 12$}. The exponents are obtained from the slope of the power spectra $\mathcal{P}_t (f)$ at radius 
  $r = 0.99 \, r_o$.
}
\label{fig:PSDalphat}
\end{figure}

The temporal power spectral density of the velocity field $\mathcal{P}_t (f) = \langle | \mathcal{F}\{\bm{u}\}(f)|^2\rangle_\phi$ is plotted in Fig.~\ref{fig:PSDt} versus the frequency $f$ for various values of the Weissenberg number in the range $12<\mathrm{Wi}<18$. Here, $\mathcal{F}\{\bm{u} \}$ denotes the temporal Fourier transform of the velocity field $\bm{u}$ and $\langle \dots \rangle_\phi$ indicates the average over the azimuthal angle, which we perform due to the angular symmetry of the geometry. We observe that the 
fluctuations in the fluid flow are strongest near the edge of the outer cylinder and in the midplane of the parallel plate geometry. 
Therefore, the power spectra are taken at radius $r=0.99\, r_o$ and height $z=0.45\,d$.

All the energy spectra in Fig.\ \ref{fig:PSDt} are plotted for Weissenberg numbers above $\mathrm{Wi}_c$. We observe a gradual 
decrease at low $f$ and a steep decline between $f \approx 10$ and $f \approx 100$. It can be fitted by a power-law decay $\mathcal{P}_t(f) \propto f^{-\alpha_t}$ as the dashed lines indicate. The power-law spectrum hints at a randomly fluctuating velocity flow field in time, where most temporal frequencies are excited and correlations decrease fast.
Fitting the temporal exponent $\alpha_t$, we observe $2.99\leq \alpha_t \leq 3.30$ for all $\mathrm{Wi}\leq 16.34$. 
Our largest value for $\alpha_t$ is comparable to the values of $\alpha_t = 3.5$ and $3.6$ found in experiments on the parallel plate geometry \cite{groisman2000,groisman2004elastic}. 

Furthermore, the characteristic exponent $\alpha_t$ also depends on the height $z$.
We show this dependence for different \Wi in Fig.~\ref{fig:PSDalphat}. 
A clear increase of $\alpha_t$ towards the midplane in the two-plate geometry is observed due to increased velocity fluctuations. Additionally, $\alpha_t$ shows non-monotonic behavior
in \Wi. It first increases after the transition at $\mathrm{Wi}_c$ and {then decreases for} $\mathrm{Wi} > 14.44$.

\subsubsection{Spatial}

\begin{figure}
\centering
\includegraphics[width=0.48\textwidth]{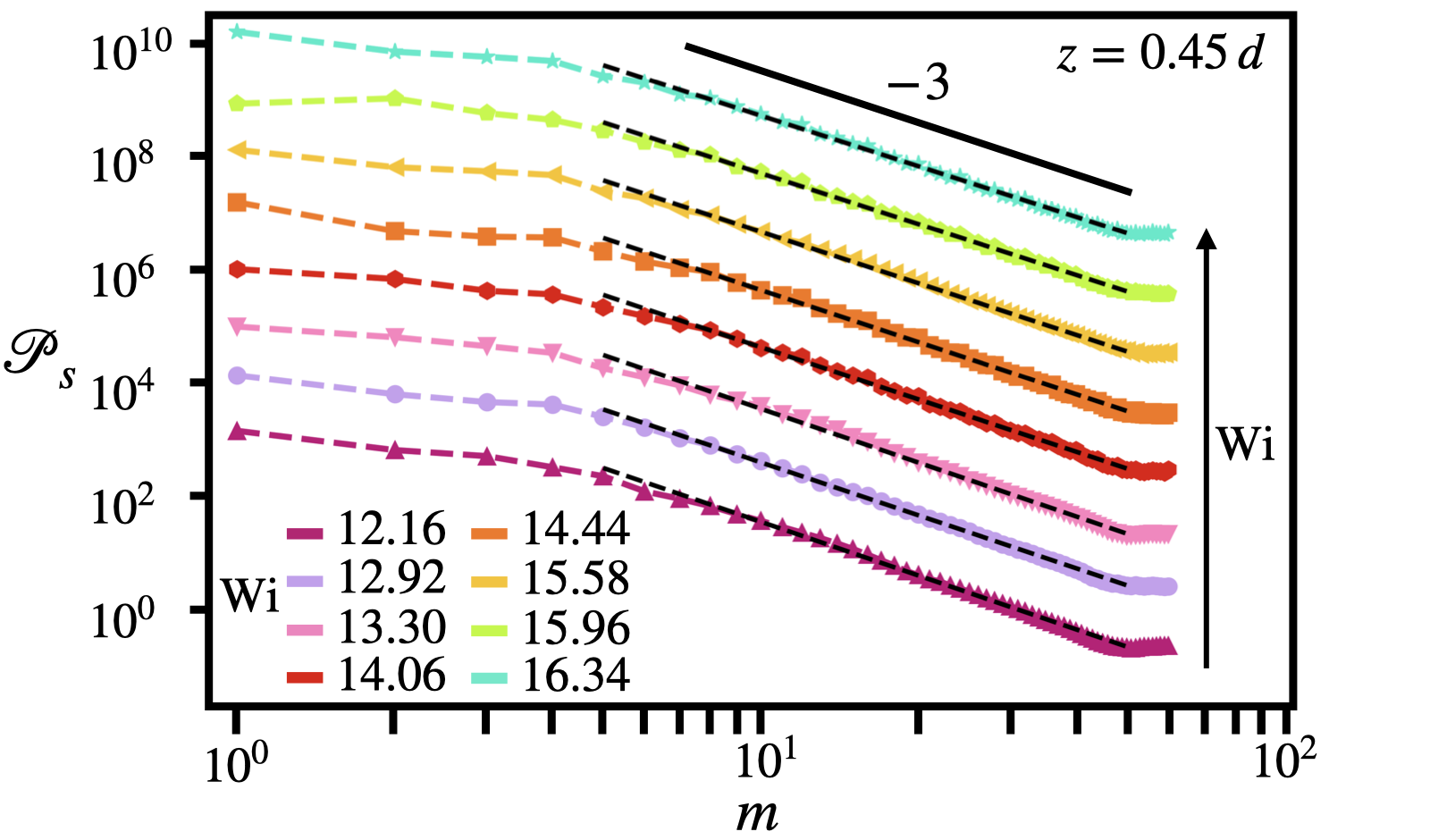}
  \caption{Spatial power spectral density 
  {$\mathcal{P}_s (m) = \mean{ |\mathcal{F}\{ \bm{u} \}(m) |^2}$} of the velocity fluctuations as a function of the azimuthal wave number $m$ for different \Wi. The spectra are taken
at height $z=0.45 \,d$ and radial position $r=0.99\,r_o$.
The data are arbitrarily offset for clarity by multiplying with a constant.
Power laws $m^{-\alpha_s}$ are fitted and plotted as black dashed lines in the range
$ 5 \leq m \leq 50$.
}
\label{fig:PSDs}
\end{figure}
\begin{figure}
\centering
\includegraphics[width=0.41\textwidth]{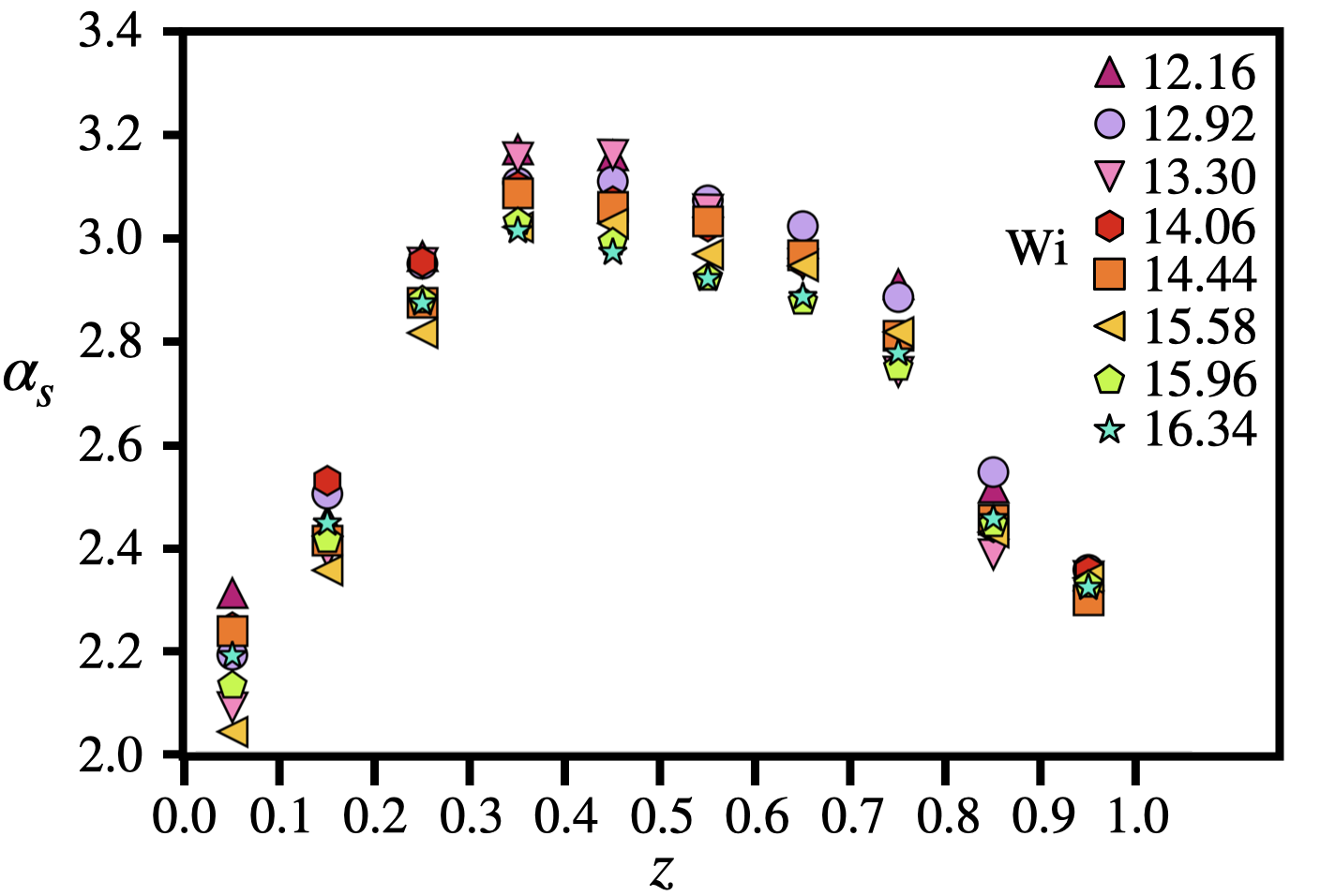}
  \caption{
  Spatial characteristic exponent $\alpha_s$ plotted versus the height $z$ for several values of the Weissenberg number $\mathrm{Wi}$ above 
  $\mathrm{Wi}_c = 12$. The exponents are obtained from the slope of the power spectra $\mathcal{P}_s (m)$ at radius $r = 0.99 \, r_o$.
 }
\label{fig:PSDalphas}
\end{figure}

In Fig.\ \ref{fig:PSDs} we plot the spatial power spectra of the velocity field 
$\mathcal{P}_s (m) = \mean{ |\mathcal{F}\{ \bm{u} \}(m) |^2}$ along the azimuthal direction against the azimuthal wave number $m$ for several $\mathrm{Wi} > \mathrm{Wi}_c$.
Again, we determine the spectra at radius $r=0.99\, r_o$ and height $z=0.45 \, d$ and the bar indicates the temporal average.
We observe a monotonic decrease in energy density with increasing azimuthal mode $m$ in all power spectra. The energy located in the first four modes is nearly constant while beyond $m=4$ it sharply decreases with $m$ according to the power law $\mathcal{P}_s(m)\sim m^{-\alpha_s}$, where $\alpha_s$ is the spatial exponent.
The dashed lines are fits in order to determine $\alpha_s$. 
In Fig.~\ref{fig:PSDalphas} we plot the spatial exponent $\alpha_s$ versus height $z$ for different \Wi.
The exponent $\alpha_s$ shows similar behavior as its temporal counterpart. It is smallest near the upper and lower plate and largest in the midplane of the geometry, for all Weissenberg numbers above the transition. 
This behavior is again consistent with our observation that the largest velocity fluctuations occur in the midplane. For
$0.3<z<0.7$, the exponent $\alpha_s$ is larger than 3 in agreement with
the theoretical condition for elastic turbulence derived in Ref.~\onlinecite{fouxon2003spectra}.
Our maximum value of $\alpha_s = 3.16$ at $z=0.45$ and $\mathrm{Wi} = 12.16$ or $\mathrm{Wi} = 13.30$ is again close 
to the experimentally observed value of $\alpha_s = 3.5$ \cite{burghelea2007elastic}.

The overall trend of both temporal and spatial exponents is similar as well as the maximum values of both exponents. However, 
the spread of values in the temporal exponent is much larger than in the spatial exponents, which are all grouped closely together, 
indicating that the spatial exponents are more accurate. Our results suggest that  Taylor's hypothesis for inertial turbulence, 
according to which both exponents should be equal \cite{taylor1938spectrum}, is not completely fulfilled in our simulations.
However, the difference between the exponents is small and the behavior of both exponents as a function of the height is remarkably 
close, which implies that also applying Taylor's hypothesis to elastic turbulence is often reasonable.

\rev{
\section{Discussion: Comparison to experiments}
\label{sec.discussion}
}

\rev{
In our work, we did not aim to quantitatively reproduce the results of a specific experiment
on the van K\'arm\'an swirling flow. Instead, due to the numerical complexity we first present an investigation of the general features of the elastic instability and elastic turbulence, as we observe them in our simulations. In the following we discuss our findings also in the context of experimental results.\\

\subsection{Geometric parameters}

We first note that even a dilute polymer solution can exhibit complex behavior, which is not captured by the \mbox{Oldroyd-B} model, an idealized polymeric fluid model. 
This includes  shear thinning, viscous heating, polymer degradation, finite polymer extensibility, and a broad spectrum of relaxation times. Thus a detailed comparison with experiments is hard at this stage. We therefore decided to use the \mbox{Oldroyd-B} model to capture main features of  the elastic instability and elastic turbulence.

Moreover, we implemented a lateral boundary condition, where we allow tangential flow, while in experiments lateral walls with 
no-slip boundary conditions are used to constrain the fluid. There, the distance between the rotating plate and the lateral wall is an important parameter, especially when the distance becomes small. This creates high shear gradients  in the corner between the 
rotating plate and the static wall. As a result, a secondary flow with curved streamlines can arise such as in the case of the Taylor-Couette geometry \cite{davoodi2018secondary}.
Our aim was to avoid such complications.

Finally, the aspect ratio of plate radius to plate separation influences the transitional path to elastic turbulence \cite{mckinley1991observations}.
We chose a large value to achieve a more uniform shear rate along the gap between the plates and to obtain a linear shear profile in the radial direction. Since the largest shear rate occurs at the lateral sides of the swirling flow, the elastic instability first occurs there.

\subsection{Onset of the instability}
In experiments radially localized disturbances become visible at the critical Weissenberg number \cite{mckinley1991observations,schiamberg2006transitional}. They are described
by the linear stability analysis of Ref.~\onlinecite{oztekin1993instability}, which shows that the critical Weissenberg number and the azimuthal wave number $m$ of the most unstable mode depends on different parameters such as the gap-to-plate-radius ratio, Deborah number, and the viscosity ratio $\beta$. Furthermore, experiments show that depending on the different parameters, the elastic instability can be either continuous or hysteretic \cite{schiamberg2006transitional,burghelea2007elastic}.

Following the approach of Ref.\ \onlinecite{oztekin1993instability}, we derived a corrected set of equations, from which we identified the non-axisymmetric mode with three-fold symmetry to be the most unstable at the critical value $\mathrm{Wi_c^{LSA}} = 11.255$.
Other non-axisymmmetric modes with two- and four-fold symmetries and the axisymmetric mode become unstable slightly above $\mathrm{Wi} = 12$, all with radial wave numbers $\alpha$ between 3.25 and 3.5. Shallow minima in the corresponding neutrality curves in the $\alpha - \mathrm{Wi}$ plane indicate that many modes with similar $\alpha$ become unstable at the same Weissenberg number.
In contrast, in our simulations near the transition, we observe an unstable non-axisymmetric mode with four-fold symmetry, which drives the laminar flow towards weakly chaotic and turbulent flow. It develops at the outer edge of the swirling flow, where the shear rate is the highest. This observation agrees well with experiments \cite{mckinley1991observations,schiamberg2006transitional} and corresponds to the localized radial position assumed in the linear stability analysis. The discrepancy between simulations and the linear stability analysis is most likely due to the fact that the latter uses spatial decoupling in the ansatz for the flow field and considers flow between infinitely extended parallel plates.

Our subcritical transition displays hysteretic behavior, which has also been observed in experiments \cite{groisman2000,groisman2004elastic,burghelea2007elastic,mckinley1991observations}.
When we now lower $\mathrm{Wi}$ below the critical value $\mathrm{Wi_c}  = 12$, the remaining weakly chaotic flow fluctuates around a three-fold symmetric pattern ($|m| = 3$). Ultimately, further lowering $\mathrm{Wi}$ below $\mathrm{Wi_c^{LSA}}$ we observe a clear non-axisymmetric mode, which decays to zero, in good agreement with our linear stability analysis. What we do not understand is why the $|m| = 3$ mode does not develop from the base flow, when we increase $\mathrm{Wi}$. A possible reason is that this requires much longer simulations, which we cannot perform. Note also in the experiments of Ref.\ \onlinecite{magda1988transition} a long duration of ca. $2100\, \mathrm{s} \approx 100 \,\lambda$ was needed to observe the elastic instability.

\subsection{Bistable flow state and order parameter}
In experiments different transition paths are observed beyond the critical Weissenberg number. Either the first transition is subcritical into a non-axisymmetric single mode followed by elastic turbulence \cite{burghelea2007elastic} or a sequence of continuous transitions between periodic and aperiodic modes occurs before elastic turbulence is entered 
\cite{schiamberg2006transitional}. In constrast, in our simulations above the critical Weissenberg number, a bistable flow state emerges, which has not been reported so far in experiments. The weakly chaotic flow displays small velocity fluctuations around a small mean value $\langle\sigma\rangle \approx 0.02$ and the turbulent flow shows large velocity fluctuations around a larger mean value of $\langle\sigma\rangle \approx 0.1$. When \Wi is closer to $\mathrm{Wi}_c$, the bistable flow state needs more time to develop from the laminar base flow. 
This was also observed in experiments\cite{mckinley1991observations}.

\subsection{Flow resistance}
In accordance with the subcritical transition from laminar base flow to the non-axisymmetric mode, the experiments in Ref.\ \onlinecite{burghelea2007elastic} also observe a sharp increase in the flow resistance and hysteresis. This agrees with the numerical findings in our work.  The flow resistance sharply increases at the transition by a factor of 1.4 and shows hysteresis. The increase is mainly due to the turbulent flow and work performed on the open side surface of the swirling flow, while the contribution from the weakly chaotic flow gradually increases starting at $\mathrm{Wi}_c$. Particularly, the elastic part of the work outweighs the viscous part, which demonstrates the elastic nature of the transition. Note that in experiments performed on swirling flows with a fixed outer boundary \cite{groisman2000,groisman2001efficient,groisman2001stretching,groisman2004elastic,burghelea2007elastic,jun2009power,jun2017polymer,schiamberg2006transitional} all work is solely performed on the top surface of the fluid and the increase in the flow resistance is directly measurable by the work required to rotate the top plate. Finally, in experiments in the regime of elastic turbulence the flow resistance shows power law scaling with an exponent of 0.49 \cite{burghelea2007elastic}, while we only observe a small increase.

\subsection{Power spectrum spatial and temporal}
In experiments the temporal power spectra of the velocity fluctuations are measured in the center of the swirling flow \cite{groisman2000,groisman2004elastic,burghelea2007elastic}, where velocity fluctuations are assumed to be homogeneous and isotropic.
Our spatial power spectra along the azimuthal direction can only be determined outside the center for $r \ne 0$. Since in our case velocity fluctuations are largest at the open side surface of the swirling flow, we decided to record the spectra near this location. While the velocity fluctuations become anisotropic, we do not expect the scaling exponents to strongly change since we always use the magnitude of the Fourier transformed velocity vector, when determining the spectra.

In the turbulent flow state, similar to experiments, we also observe power-law scaling in the spatial and temporal power spectral densities.
The power spectra were analyzed at different heights $z$ between the two plates.
The steep decay observed in the spatial power spectra shows that the magnitude of velocity fluctuations occurring at large spatial modes is greatly reduced and velocity fluctuations mainly occur on large spatial scales (on the order of the system size). 
Hence, the velocity field is spatially smooth, dominated by strong nonlinear interactions of a few large-scale spatial modes \cite{steinberg2021elastic}.
In the height range $0.3< z <0.7$, we find the spatial scaling exponent $\alpha_s$ in $\mathcal{P}_s(m)\sim m^{-\alpha_s}$ to be above 3, which is a determining feature of elastic turbulence \cite{fouxon2003spectra}.  Both, spatial and temporal exponents show similar behavior as a function of the height $z$ and are largest in the midplane, where velocity fluctuations are largest. The maximum values of the temporal ($\alpha_t = 3.41$) and spatial ($\alpha_s= 3.17$) exponents in our simulations are remarkably close to the experimentally observed values \cite{groisman2000,groisman2004elastic,burghelea2007elastic}. Since both exponents show the same behavior as a function of the height $z$ and their values are similar, our results suggest that applying Taylor's hypothesis to numerical solutions displaying elastic turbulence is usually reasonable.
}

\rev{
\section{Conclusion}
\label{sec:conclusion}
In summary, we have presented a detailed analysis of the three-dimensional von K\'arm\'an swirling flow between two parallel plates through numerical solutions of the \mbox{Oldroyd-B} model. 
We characterized the flow state with the secondary-flow strength $\sigma$, a measure of the average strength of the velocity fluctuations, and then defined an order parameter $\Phi$ as the time average of $\sigma$.
At the critical Weissenberg number $\mathrm{Wi_c}=12$, the order parameter displays a subcritical transition from laminar flow to a bistable flow, which switches between
\rev{weakly chaotic flow}, 
where $0.01 < \Phi < 0.06 $, and elastic turbulence with $\Phi > 0.07$.
We defined the flow resistance as the total work performed on the fluid and found a significant increase of the flow resistance at the transition, due to the elastic part of the fluid in the turbulent flow state.
In addition, we observed hysteretic behavior in both the secondary-flow strength and the flow resistance, which is a common feature of a subcritical transition and has also been found in experiments \cite{groisman2000,groisman2004elastic,burghelea2007elastic,mckinley1991observations}. 
Furthermore, we analyzed the spatial and temporal power spectra of the velocity field. Their power-law decays reflect the turbulent nature of the flow. Overall, we observe a different transitional route to elastic turbulence than in the experiments of Groisman and Steinberg \cite{groisman2000,groisman2004elastic}, Burghelea \textit{et al.} \cite{burghelea2007elastic} and  Schiamberg \textit{et al.} \cite{schiamberg2006transitional}. However, the observed properties of elastic turbulence remain similar.

Our numerical simulation study helps to further understand the transition from laminar to turbulent flow and to characterize viscoelastic fluid flow beyond the transition. In particular, 
we are able to monitor physical quantities such as the elastic stress fields, which are not directly accessible in experiments. However, polymer solutions used in the experiments have more complex behavior than the relatively simple \mbox{Oldroyd-B} model.  
Several directions exist to extend our investigations.
More detailed numerical work is recommended to investigate the elastic instability near the transition from laminar to chaotic or turbulent flow. In particular, studies involving varying the ratio of plate radius to plate distance and the polymer-to-solvent viscosity ratio are warranted. Also, more refined polymer models that include shear thinning and/or finite polymer extensibility, such as the FENE-P model, should be studied.
However, detailed parameter studies are still limited due to the large computational effort and the large amount of data storage required. 
Moreover, the influence of different boundary conditions of the swirling flow at its lateral side has to be analyzed further, both numerically and experimentally.
Also, a nonlinear stability analysis of the instability, which has not yet been performed, could further illuminate the transition.
One important characteristic of elastic turbulence, the interesting application of mixing solutes by elastic turbulent flow, has not been inspected in simulations, yet.
We aim to address some of these open questions in future work.
}

\begin{acknowledgments}
We acknowledge support from the Deutsche Forschungsgemeinschaft in the framework of the Collaborative Research  Center SFB 910.
The work was supported by the North-German Supercomputing Alliance (HLRN). We are grateful to the HLRN supercomputer staff for support.
\end{acknowledgments}

\section*{Author Contributions}
All authors contributed equally to this work.

\section*{Conflict of Interest}
The authors have no conflicts to disclose.

\section*{Data Availability Statement}
The data that support the findings of this study are available from the corresponding author upon reasonable request.

\appendix

\section{Spatial mesh}

\begin{figure}
\rev{
\centering
\includegraphics[width=0.41\textwidth]{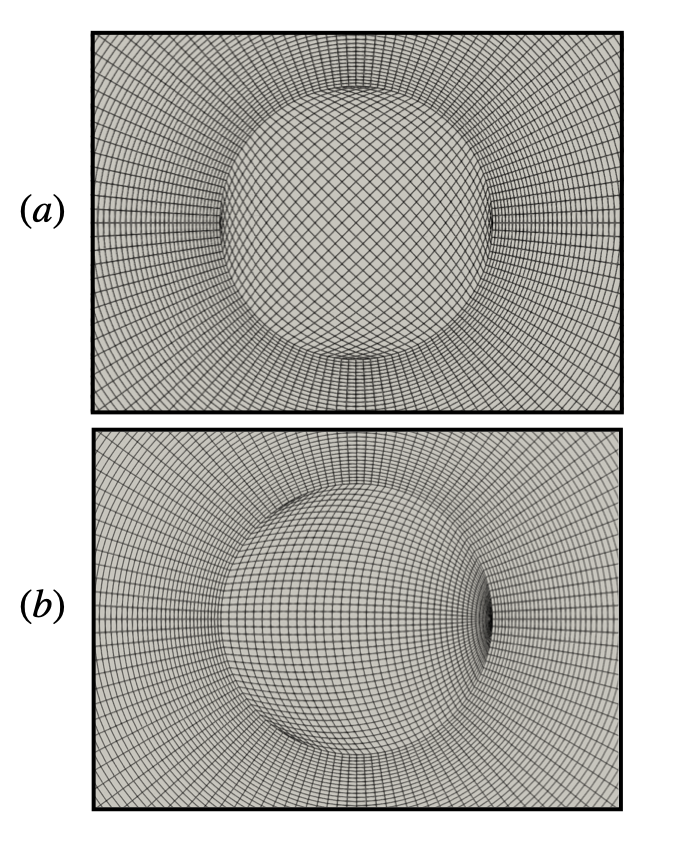}
  \caption{ 
  Overview of the inner cylinder block comprising a square-like lattice employed in the simulations. The faces are connected to the mesh of the the outer cylinder, which is constructed from (a) four connected trapezoid blocks and (b)  three connected trapezoid blocks.
}
\label{fig:Mesh}
}
\end{figure}

\rev{Here we give a brief description of the special square-like mesh employed in the inner cylindrical block to avoid the singularity at $r=0$. The internal matching of the cells leads to small mesh volumes (visible by dark regions in Fig.~\ref{fig:Mesh}),
which in turn lead to small numerical errors on the order of $0.0013 \, u^0_\mathrm{max}$, where $u^0_\mathrm{max} = r_o \Omega$ is the maximum base flow velocity. The numerical errors have the same spatial symmetry as the symmetry of the connected trapezoid blocks comprising the outer cylinder. To check the accuracy of the four-fold azimuthal pattern developing (see Sec.~\ref{sec:onset}), we performed simulations with the outer cylinder consisting out of four and out of three connected blocks, see Fig.~\ref{fig:Mesh}. For both meshes we find a four-fold symmetric unstable mode. 
}

\section{Initial disturbance wave}
\begin{figure*}
\rev{
\centering
\includegraphics[width=0.95\textwidth]{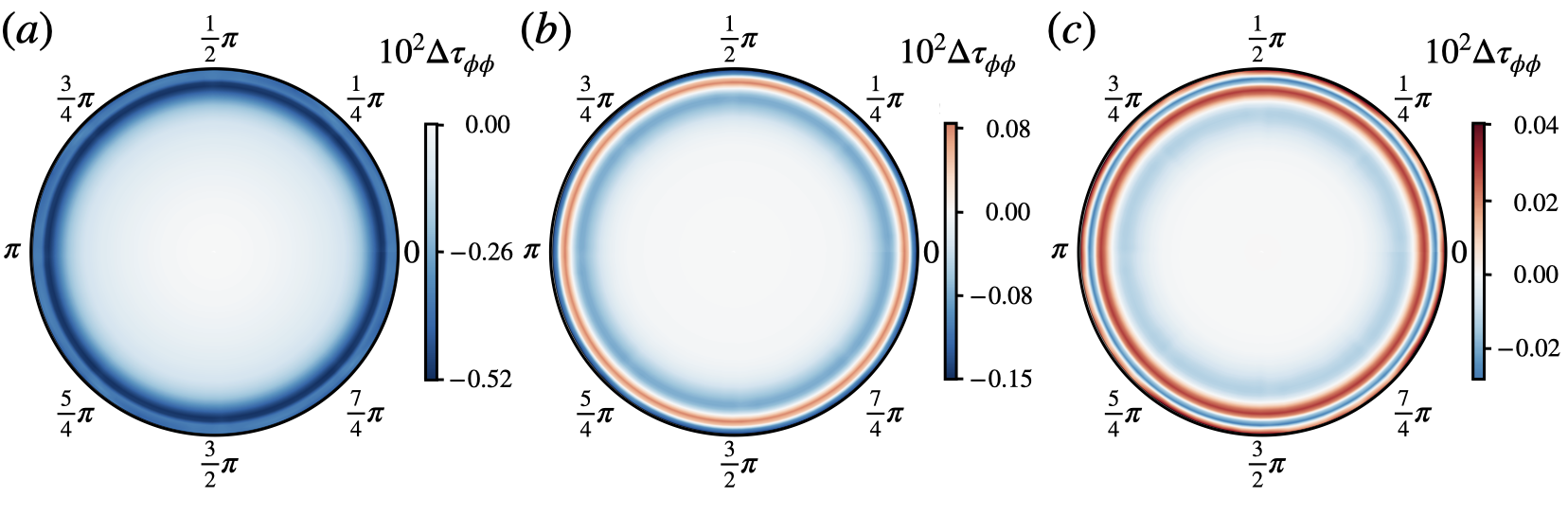}
  \caption{ 
    (Multimedia view) Azimuthal stress-tensor component color-coded in the $r-\phi$ plane near the mid plane  $z=0.45\,d$ at $\mathrm{Wi}=12.008$. The component $\Delta \tau_{\phi\phi} = \tau_{\phi\phi}-\overline{\tau_{\phi\phi}}|_{t_s}^{t_e}$ is given relative to the mean value between times $t_s=2$ and $t_e=12$. Snapshots of the stress field are taken at times: \mbox{(a) $t=1.2$}, \mbox{(b) $t=1.85$} and \mbox{(c) $t=2.44$}.
    An axisymmetric disturbance is observed, which travels inwards and decays over time.
}
\label{fig:disturbance}
}
\end{figure*}

\rev{
Here we display the onset of the instability from a different perspective. The onset is equivalent to the one described in Sec.~\ref{sec:onset}. We first observe an initial axisymmetric disturbance flow, which begins at the outer edge and then travels inwards, see Fig.~\ref{fig:disturbance}. In the figure, the axisysmmetric elastic wave can be seen traveling inwards at the outer edge with snapshots given in Fig.~\ref{fig:disturbance} (a)-(c) with increasing time.
Afterwards, an unstable non-axisymmetric mode with azimuthal wave numbers $m= \pm 4$ develops which drives the flow towards weakly chaotic flow 
(shown in the video) and ultimately towards elastic turbulence (not shown in the video).
}

\begin{widetext}
\section{Linear Stability Analysis}
\label{app1}

In our linear stability analysis we perform a classical normal-mode perturbation, following the work of Ref.~\onlinecite{avgousti1993non} and \rev{\onlinecite{oztekin1993instability}}.
We decouple the spatial dependencies of the perturbation and superimpose an infinitesimal disturbance ($ \bm{\tau}', \bm{u}',p'$) on the base flow solution ($ \bm{\tau} ^0, \bm{u}^0,p^0$). We further consider an infinitely extended flow in the radial direction, which is bounded between two plates in the z-direction.
The perturbation is assumed to occur at a critical radius $r_c$\rev{, where the Weissenberg number reaches its critical value: $\mathrm{Wi}_c=\lambda \Omega r_c/d$}.
We set the top plate at $z=1$ fixed and the lower plate at $z=0$ to be moving with angular velocity $\Omega$.
Neglecting fluid inertia ($\text{Re} \approx 0$), the steady-state velocity field
for infinitely extended parallel plates is purely azimuthal and given by
\begin{align}
\label{Aeq:baseflow}
    \bm{u}^0 = \frac{r\Omega }{ d }(1-z) \bm{e}_\phi \, .
\end{align}
The corresponding base stress and pressure fields read
\begin{align}
\label{Aeq:taubase}
   & \tau_{rr}^0 = \tau_{r\phi}^0 = \tau_{r z}^0 = \tau_{zz}^0 = 0; \qquad \tau_{\phi z}^0 = -\eta_p \frac{r\Omega}{d} ; \\
   & \tau_{\phi\phi}^0 = 2 \eta_p \lambda \left(\frac{r\Omega}{d}\right)^2 ; \qquad
    p^0 = -\eta_p \lambda \left(\frac{r\Omega}{d}\right)^2 \, .
\end{align}
The velocity, stress and pressure fields are then expressed as
\begin{align}
\label{eq:linstabbase}
\tilde{\bm{\tau}} =  \bm{\tau} ^0 + \bm{\tau}'  \, ;\qquad
\bm{\tilde u} =   \bm{u}^0 + \bm{u}' \, ;\qquad\tilde p = p^0 + p'  \, .
\end{align}

After substituting Eq. (\ref{eq:linstabbase}) in the governing equations (\ref{eq:NS})-(\ref{eq:Oldroyd}), subtracting the base flow given by the velocity field Eq. (\ref{Aeq:baseflow}), stress and pressure fields Eq. (\ref{Aeq:taubase}) and keeping only terms that are linear in the disturbance functions, we obtain a set of 10 ordinary differential equations described in the following.
Dimensionless variables are obtained by normalizing the length by $d$, the time by $\Omega^{-1}$, the velocity by $d\Omega$ and the stress by $(\eta_s + \eta_p) \Omega$. We denote the dimensionless variables by $(r,\phi,z)$ and obtain for the continuity equation,
\begin{align}
\label{eq:AppendixFullset1}
    \frac{1}{r} \frac{\partial \, r\, u_r}{\partial r} + \frac{1}{r} \frac{\partial u_\phi}{\partial \phi}
    + \frac{\partial u_z}{\partial z} = 0 \, ,
\end{align}
the three linearized momentum equations,
\begin{align}
      \frac{1}{r} \frac{\partial }{\partial r} \left( r \tau_{rr} \right)
    + \frac{1}{r} \frac{\partial \tau_{r\phi}}{\partial \phi}
    + \frac{\partial \tau_{rz}}{\partial z}
    - \frac{\tau_{\phi\phi}}{r}
    - \frac{\partial p}{\partial r}  
    +\kappa \left( \nabla ^2 u_r - \frac{2}{r} \frac{\partial u_\phi}{\partial \phi} - \frac{u_r}{r^2} \right) = 0 \, ,
\\
     \frac{\partial \tau_{r\phi} }{\partial r}
    + \frac{1}{r} \frac{\partial \tau_{\phi\phi}}{\partial \phi}
    + \frac{\partial \tau_{\phi z}}{\partial z}
    +2 \frac{\tau_{r\phi}}{r}
    - \frac{1}{r} \frac{\partial p}{\partial \phi} 
    +\kappa \left( \nabla ^2 u_\phi + \frac{2}{r^2} \frac{\partial u_r}{\partial \phi} - \frac{u_\phi}{r} \right) = 0 \, ,
 \\
    \frac{1}{r} \frac{\partial }{\partial r} \left( r \tau_{rz} \right)
    + \frac{1}{r} \frac{\partial \tau_{\phi z}}{\partial \phi}
    + \frac{\partial \tau_{zz}}{\partial z}
    - \frac{\partial p}{\partial z} 
    +\kappa \nabla ^2 u_z = 0 \, ,
\end{align}
where $\nabla ^2$ is the Laplacian operator in cylindrical coordinates and $\kappa={\eta_s}/(\eta_s + \eta_p)$,
and six linearized stress constitutive equations
\begin{align}
    L(\tau_{rr}) &= 2 \kappa_p \frac{\partial u_r}{\partial r} \, ,\\
    L(\tau_{r\phi}) + \mathrm{De}\, r \tau_{rz} +  \kappa_p \mathrm{De}\, r \left[ \frac{\partial u_r}{\partial z} + 2\rev{\mathrm{De}} \left(u_\phi-\frac{\partial u_r}{\partial \phi} \right)  \right]
    &= \kappa_p \left( \frac{1}{r} \frac{\partial u_r}{\partial \phi} - \frac{u_\phi}{r} + \frac{\partial u_\phi}{\partial r}    \right) \, ,\\
    L(\tau_{rz}) + \kappa_p \mathrm{De}\, \left(\frac{\partial u_r}{\partial \phi} - u_\phi \right)
    &= \kappa_p \left( \frac{\partial u_r}{\partial z} + \frac{\partial u_z}{\partial z} \right) \, ,\\
    L(\tau_{\phi\phi}) + 2 \mathrm{De}\, r \tau_{\phi z} + 2 \kappa_p \mathrm{De}\, r \left[  \frac{\partial u_\phi}{\partial z} - 2 \rev{\mathrm{De}}  \left(  u_r +\frac{\partial u_\phi}{\partial \phi}\right) \right]
    &= 2\kappa_p \left( \frac{1}{r} \frac{\partial u_\phi}{\partial \phi} + \frac{u_r}{r} \right) \, ,\\
    L(\tau_{\phi z}) + \mathrm{De}\, r \tau_{zz} +\kappa_p \mathrm{De}\,r \left( \frac{\partial u_z}{\partial z} + \frac{1}{r} \frac{\partial u_\phi}{\partial \phi} +\frac{ u_r}{r}- 2 \rev{\mathrm{De}}  \frac{\partial u_z}{\partial \phi}  \right)
    &= \kappa_p \left(  \frac{1}{r}\frac{\partial u_z}{\partial \phi} + \frac{\partial u_\phi}{\partial z} \right) \, ,\\
    L(\tau_{zz}) + 2 \kappa_p \mathrm{De}\, \frac{\partial u_z}{\partial \phi}  &= 2 \kappa_p \frac{\partial u_z}{\partial z}  \, ,
    \label{eq:AppendixFullset2} 
\end{align}
where the operator $L \equiv [1 + \mathrm{De}\, \partial/\partial t + \mathrm{De}\, (1-z) \partial/\partial \phi ]$ and $\kappa_p = ( 1 - \kappa)$.
Here, we introduced the Deborah number, the ratio of the fluid relaxation time to the angular velocity of the rotating plate, $\mathrm{De} = \lambda  \Omega$. \rev{The Weissenberg number is then given by $\mathrm{Wi} = \mathrm{De} \, r_c / d$}.

Next, we insert the normal-mode perturbation and express the field equations by
\begin{align}
\label{eq:linstabNM1}
\tilde{\bm{\tau}}  &=  \bm{\tau} ^0 + \mathcal{R} \left[\bm{\tau} (z)\exp(i \alpha r + i m \phi + \sigma t)\right] \, , \\
\label{eq:linstabNM2}
\bm{\tilde u} &=   \bm{u}^0 + \mathcal{R}\left[\bm{u}(z)\exp(i \alpha r + i m \phi + \sigma t)\right] \, , \\
\label{eq:linstabNM3}
\tilde p &= p^0 +  \mathcal{R}\left[ p(z)\exp(i \alpha r + i m \phi + \sigma t) \right] \, ,
\end{align}
where $\mathcal{R}[\dots]$ represents the real part, $\bm{\tau} , \bm{u}, p$ are complex infinitesimal functions of the z coordinate, $\alpha$ is the dimensional radial wave number, $m$ is the azimuthal wave number and $\sigma$ is the (in general) complex temporal eigenvalue. The azimuthal wave number $m$ is an integer, which can be positive or negative, corresponding to non-axisymmetric disturbances or zero, corresponding to axisymmetric disturbances.

After substituting Eq. (\ref{eq:linstabNM1})-(\ref{eq:linstabNM3}) in our set of linearized equations Eq. (\ref{eq:AppendixFullset1})-(\ref{eq:AppendixFullset2}), we obtain the following set of linear disturbance equations:
\begin{align}
    \left(\frac{1}{r} + i\alpha \right)u_r + \frac{i m}{r}   u_\phi 
    + \frac{\partial u_z}{\partial z} = 0 \, , 
\end{align}
\begin{align}
     \left(\frac{1}{r} + i\alpha \right) \tau_{rr} 
    + \frac{i m}{r}  \tau_{r\phi}
    + \frac{\partial \tau_{rz}}{\partial z}
    - \frac{\tau_{\phi\phi}}{r}
    - i \alpha p
    +\kappa \left[\frac{\partial^2 u_{r}}{\partial z^2} -(\alpha^2 + m^2) u_r - \frac{2i m}{r} u_\phi - \frac{u_r}{r^2} \right] = 0 \, ,
\\
    \left(\frac{2}{r} + i\alpha \right) \tau_{r\phi}
    + \frac{i m}{r} \tau_{\phi\phi}
    + \frac{\partial \tau_{\phi z}}{\partial z}
    - \frac{i m}{r}  p 
    +\kappa \left[ \frac{\partial^2 u_{\phi}}{\partial z^2} -(\alpha^2 + m^2) u_\phi + \frac{2i m}{r^2} u_r - \frac{u_\phi}{r} \right] = 0 \, ,
\\
   \left(\frac{1}{r} + i\alpha \right)\tau_{rz} 
    + \frac{i m}{r} \tau_{\phi z}
    + \frac{\partial \tau_{zz}}{\partial z}
    - \frac{\partial p}{\partial z} 
    +\kappa \left[\frac{\partial^2 u_{z}}{\partial z^2} -(\alpha^2 + m^2) u_z \right] = 0 \, ,
\end{align}
\begin{align}
    \frac{1}{2}\mathcal{L} \tau_{rr}
    &=  i \alpha  u_r  \, ,\\
    \mathcal{L} \tau_{r\phi} +  \mathrm{De}\, r\left( \frac{\tau_{rz}}{\kappa_p} - 2 i m \mathrm{De}\,u_r +  \frac{\partial u_r}{\partial z} \right) 
    &=  \frac{i m}{r}  u_r +\left(i\alpha -\frac{1}{r} \right)u_\phi \, ,\\
    \mathcal{L} \tau_{rz} +   i m  \mathrm{De}\,   u_r 
    &=    \frac{\partial u_r}{\partial z} + \frac{\partial u_z}{\partial z}  \, ,\\
    \frac{1}{2}\mathcal{L} \tau_{\phi\phi} + \mathrm{De}\, r\left( \frac{\tau_{\phi z}}{\kappa_p} - 2 i m \mathrm{De}\,  u_\phi + \frac{\partial u_\phi}{\partial z} \right)
    &= \frac{ i m}{r} u_\phi + \frac{u_r}{r}  \, ,\\
    \mathcal{L} \tau_{\phi z}  + \mathrm{De}\,r\left( \frac{\tau_{zz}}{\kappa_p} + \frac{i m}{r} u_\phi - 2i m \mathrm{De}\,  u_z  + \frac{\partial u_z}{\partial z} \right)
    &=   \frac{i m}{r} u_z + \frac{\partial u_\phi}{\partial z} \, ,\\
    \frac{1}{2}\mathcal{L} \tau_{zz} +    i m  \mathrm{De}\,   u_z &=    \frac{\partial u_z}{\partial z}  \, ,
\end{align}
where the linear operator $\mathcal{L} \equiv \kappa_p^{-1}[1 + \mathrm{De}\, \sigma +   i m \mathrm{De}\, (1-z) ]$.

\begin{table}
\caption{Real and imaginary parts of the most unstable eigenvalue $\sigma_\mathrm{re}$ and $\sigma_\mathrm{im}$, respectively, for axisymmetric (m=0) disturbances.
The eigenvalue $\rev{\sigma^\mathrm{ref}}$ is a nonzero solution to the equation $\psi' = \rev{F^\mathrm{ref}}(z,\psi)$\rev{,} the eigenvalue $\sigma$ is a nonzero solution to the equation $\psi' = F(z,\psi)$ \rev{and $\sigma^\mathrm{lit}$ is the eigen value obtained from Ref. \onlinecite{oztekin1993instability}}.
The axial wave number $\alpha = 3.5$, the polymer to solvent viscosity ratio $\kappa =0.59, \beta = 41/59 \approx 0.695$, and the critical radius $r_c=5$. 
\rev{The number of collocation points is $N \geq 50 $.}
}
\renewcommand{\arraystretch}{1.3}
\begin{ruledtabular}
\begin{tabular}{ cccccccc } 
  &  \multicolumn{2}{l}{ Literature \cite{oztekin1993instability} } & \multicolumn{4}{l}{ This work }\\
 \hline
\rev{$\mathrm{De}$} &$\rev{\sigma_\mathrm{re}^\mathrm{lit}}$ & $\rev{\sigma_\mathrm{im}^\mathrm{lit}}$ & $\rev{\sigma_\mathrm{re}^\mathrm{ref}}$ & $\rev{\sigma_\mathrm{im}^\mathrm{ref}}$ &\rev{$\sigma_\mathrm{re}^\mathrm{ref}/\sigma_\mathrm{re}^\mathrm{lit}$} & $\sigma_\mathrm{re}$ & $\sigma_\mathrm{im}$ \\ \hline  
\rev{1} &   -0.53  &   -0.462 &   -0.50864  &   -0.44581 & \rev{0.96} &  -0.47513 & -0.48917  \\ 
\rev{2} &   -0.036 &   -0.441 &   -0.02150  &   -0.42510 & \rev{0.60} &  0.02226 &  -0.47436 \\ 
\rev{3} &   0.12  &    -0.431 &   0.13430  &   -0.41470  & \rev{1.12} &  0.18492 &  -0.46775 \\ 
\rev{5} &   0.24  &    -0.4215 &   0.25322  &   -0.40508 & \rev{1.06} &  0.30955 &  -0.45370 \\ 
\end{tabular}
\end{ruledtabular}
\label{tab:axisym}
\renewcommand{\arraystretch}{1.0}
\end{table}

\begin{table}
\caption{
Real and imaginary parts of the most unstable eigenvalue $\sigma_\mathrm{re}$ and $\sigma_\mathrm{im}$, respectively, for nonaxisymmetric ($m=1,2,3$) disturbances.
The eigenvalue $\rev{\sigma^\mathrm{ref}}$ is a nonzero solution to the equation $\psi' = \rev{F^\mathrm{ref}}(z,\psi)$, the eigenvalue $\sigma$ is a nonzero solution to the equation $\psi' = F(z,\psi)$ \rev{and $\sigma^\mathrm{lit}$ is the eigen value obtained from Ref. \onlinecite{oztekin1993instability}}. The Weissenberg number $\mathrm{Wi}=30$, the axial wave number $\alpha = 3.5$, the polymer to solvent viscosity ratio $\kappa =0.59, \beta = 41/59 \approx 0.695$, and the critical radius $r_c=5$.
\rev{The number of collocation points is $N \geq 50 $.}
}
\renewcommand{\arraystretch}{1.3}
\begin{ruledtabular}
\begin{tabular}{ cccccccc } 
  &  \multicolumn{2}{l}{ Literature \cite{oztekin1993instability} } & \multicolumn{4}{l}{ This work }\\
 \hline
m &$\rev{\sigma_\mathrm{re}^\mathrm{lit}}$ & $\rev{\sigma_\mathrm{im}^\mathrm{lit}}$ & $\rev{\sigma_\mathrm{re}^\mathrm{ref}}$ & $\rev{\sigma_\mathrm{im}^\mathrm{ref}}$ & \rev{$\sigma_\mathrm{re}^\mathrm{ref}/\sigma_\mathrm{re}^\mathrm{lit}$} & $\sigma_\mathrm{re}$ & $\sigma_\mathrm{im}$  \\ \hline  
1 & 0.13 & -1.25  &  0.13123 & -0.77516 & \rev{1.01} & 0.11270 & -0.74802 \\ 
2 & 0.17 & -0.375 &  0.15170 & -1.64974 & \rev{0.89} & 0.24631 & -1.47938 \\ 
3 & 0.09 & -0.345 &  0.07242 & -2.60319 & \rev{0.81} & -0.07971 & -2.61053 \\ 
\end{tabular}
\label{tab:nonaxisym}
\end{ruledtabular}

\renewcommand{\arraystretch}{1.0}
\end{table}

The set of linearized governing equations can be reduced from 10 independent variables to a set of 6 ordinary differential equations and 4 linear relationships.
which we express with the vector 
\begin{align}
    \rev{
    \bm{\psi} = [u_r(z),u_\phi(z),u_z(z), p(z),u_r'(z),u_\phi'(z) ]^\mathrm{T} \, ,
    }
\end{align}
where the prime indicates the derivative $g'=dg/dz$ with respect to the spatial coordinate $z$. The boundary conditions of the perturbation 
are set to
\begin{align}
    \bm{u}(r,\phi,z) = 
    {\bm{0}}
    \qquad \text{at $z = 0$ and $z = 1$} \, .
\end{align}
Moreover, we take the radially localized perturbation to occur at the edge of the rotating cylinder, $r_c=r_o$, where the shear rate is the highest. 
Next, we numerically solve the linear stability problem using a collocation method, which discretizes the perturbation in the z direction \rev{with $N$ collocation points}, with the computer algorithm \textit{solve\_bvp} available in the SciPy package \cite{2020SciPy-NMeth}.

Next, we investigate the stability of the base flow with respect to axisymmetric disturbances ($m=0$) and to nonaxisymmetric disturbances
 ($0 < m \leq 4$). First we validate our {numerical algorithm}
 and reproduce results of Ref.~\onlinecite{oztekin1993instability}, which is essential given the large number of terms involved in the calculations 
 and our new approach to numerically solving them.
In our derivation of the linearized set of equations, given by Eq. (\ref{eq:AppendixFullset1}) - (\ref{eq:AppendixFullset2}), we found extra terms compared to Ref.~\onlinecite{oztekin1993instability}. \rev{We observe two differences: in Eq. (\ref{eq:AppendixFullset1}) the ${\partial u_r}/{\partial \phi}$ term scales with $\mathrm{De}^2$ instead of $\mathrm{De}$; and in Eqs. (\ref{eq:AppendixFullset1}) - (\ref{eq:AppendixFullset2}) we obtain terms involving $u_r$ and $u_\phi$, which are missing in Ref.~\onlinecite{oztekin1993instability}.}
Therefore, we will compare the results of both sets of equations.
In the validation procedure we have tested the algorithm, where we solve the system of equations of Ref.~\onlinecite{oztekin1993instability} 
defined as $\bm{\psi}' = \rev{F^\mathrm{ref}}(z,\bm{\psi})$ to obtain the most unstable eigenvalue $\rev{\sigma^\mathrm{ref}}$ for a specific Weissenberg number. 
We set the axial wave number to $\alpha = 3.5$, the polymer to solvent viscosity ratio $\beta = 41/59 \approx 0.695$ (the polymer to total viscosity of the polymer solution ratio is $\kappa = 0.4$), and the critical radius to $r_c=5$, which are the parameters used in Ref.~\onlinecite{oztekin1993instability}.
The results for the axisymmetric case are presented in Table \ref{tab:axisym}.
The table shows good agreement between our algorithm and the results of Ref.~\onlinecite{oztekin1993instability} for axisymmetric instabilities. Moreover, the most unstable eigenvalue of the newly derived set of equations $\sigma$ is close to the most unstable eigenvalue $\rev{\sigma^\mathrm{ref}}$ of the former set of equations used in Ref.~\onlinecite{oztekin1993instability}. 

In addition, the validation results for three nonaxisymmetric cases ($m=1,2,3$) are presented in Table \ref{tab:nonaxisym} at Weissenberg number \Wi = 3. Again, we see a good agreement between our algorithm and the results of Ref.~\onlinecite{oztekin1993instability}, where the deviation in the real part of sigma is at most 0.02 and the deviation in the imaginary part is about $0.5,1.3$, and $2.3$ for $m=1,2,3$, respectively. Thus, only for the imaginary part of the most unstable eigenvalue do we observe significant deviation. However, 
${\sigma_\mathrm{im}}$ plays no role in the stability of the eigenvalue.

\end{widetext}

\bibliography{literature}

\end{document}